\documentclass[twocolumn,prb,amsmath,amssymb,superscriptaddress]{revtex4}
\usepackage[colorlinks]{hyperref}
\usepackage{xcolor}
\usepackage[utf8]{inputenc}
\usepackage{amsmath}
\usepackage{amsfonts}
\usepackage{amssymb}
\usepackage{graphicx}
\usepackage{wrapfig}
\usepackage{MnSymbol}
\usepackage{braket}
\usepackage{comment}
\usepackage{array}
\usepackage{subfigure}
\usepackage{float}
\usepackage{bm}
\usepackage{dsfont}
\usepackage{cellspace}

\DeclareSymbolFont{matha}{OML}{txmi}{m}{it}
\newcolumntype{P}[1]{>{\centering\arraybackslash}p{#1}}
\newcommand{\lv}[1]{\reflectbox{\ensuremath{\vec{\reflectbox{\ensuremath{#1}}}}}}

\allowdisplaybreaks

\begin{document}
\title{Twisted bilayer graphene at charge neutrality: \\competing orders of SU(4) Dirac fermions}

\author{Nikolaos Parthenios}
\affiliation{Max Planck Institute for Solid State Research, D-70569 Stuttgart, Germany}

\author{Laura Classen}
\affiliation{Max Planck Institute for Solid State Research, D-70569 Stuttgart, Germany}
\affiliation{Department of Physics, Technical University of Munich, D-85748 Garching, Germany}

\begin{abstract}
We study possible patterns for spontaneous symmetry breaking in a Dirac fermion model, which is applicable to twisted bilayer graphene at charge neutrality.
We show how a chiral SU(4) symmetry emerges 
and construct the corresponding low-energy model that includes  
a Fierz-complete set of symmetry-allowed four-fermion interactions. We employ an unbiased renormalization group treatment to identify the critical points that describe transitions into 
different ordered phases. The resulting phase diagram depends on the number of fermion flavours and we show that the coupling between ordering channels prevents many of the possible mean-field orders from being accessible at relevant, small flavour numbers. We argue that, as a consequence,  
twisted bilayer graphene 
is governed by a quantum Hall state or an SU(4) manifold 
 of insulating spin-valley orders  
with emergent Lorentz symmetry that contains inter-valley coherent, spin Hall, and valley Hall states. We study how SU(4)-breaking perturbations affect the accessibility and 
 can additionally stabilize symmetry-broken (semi-)metallic states.
\end{abstract}

\maketitle

\section{Introduction}

Spontaneous symmetry breaking of Dirac fermions plays an important role in many systems, ranging from the chiral phase transition in quantum chromodynamics to quantum critical points in semimetals \cite{doi:10.1080/00018732.2014.927109,doi:10.1146/annurev-conmatphys-031113-133841,Boyack2021,Braun_2012}. 
Interactions must be sufficiently strong for such phase transitions to occur because the density of states of massless Dirac fermions vanishes at charge neutrality. 
Furthermore, if the phase transition takes place at zero temperature and is continuous, fermionic quantum critical behavior is expected which does not possess any classical analogue \cite{Boyack2021}. 

As a well-known example, spontaneous symmetry breaking in graphene was intensely studied. However, interactions in graphene are estimated to be slightly too small to induce a phase transition\cite{PhysRevLett.111.056801}. 

The recent discovery of strongly correlated moir\'e materials\cite{Cao2018Corr,Cao2018Super} provides a new opportunity for the investigation of quantum phase transitions in two-dimensional Dirac systems, where the relative interaction can be tuned via a twist angle\cite{moiremarvels,Kennes2021,https://doi.org/10.48550/arxiv.2201.09260}. 
This includes, for example, symmetrically twisted few-layer graphene systems \cite{PhysRevB.100.085109}, or twisted $\Gamma$-valley transition metal dichalcogenides \cite{doi:10.1073/pnas.2021826118,XianNatComm2021}. 
In particular, the most prominent moir\'e material - twisted bilayer graphene (TBG) - also hosts Dirac fermions at charge neutrality. Although the Dirac velocity vanishes in TBG at a so-called magic angle, strong interactions are expected in its vicinity where Dirac fermions remain present \cite{doi:10.1073/pnas.1108174108,PhysRevLett.99.256802}. 

Experiments on TBG report both insulating\cite{Lu2019,doi:10.1126/science.aaw3780,doi:10.1126/science.aay5533,Stepanov2020,Pierce2021,PhysRevLett.127.197701} 

and semimetallic\cite{Cao2018Corr,Cao2018Super,doi:10.1126/science.aav1910,Park2021,Zondiner2020,doi:10.1126/science.abb8754,Uri2020,Rozen2021} behavior near charge neutrality with the differences potentially coming from twist-angle disorder, strain, or substrate alignment. In particular, ultra-low strain samples show the emergence of a gap \cite{nuckolls2023quantum}. Theoretically, several ground states were proposed for TBG near charge neutrality. \cite{PhysRevX.8.031089,Bultinck2020Aug,PhysRevLett.124.097601,Liu2021Jan,PhysRevX.11.011014,PhysRevX.12.011061,PhysRevResearch.4.033168}
It was argued that different ordered states lie close in energy because the interaction of the continuum model for TBG possesses a U(4) symmetry, or even U(4)$\times$U(4) in the chiral limit, which relates the ordered states\cite{Bultinck2020Aug,PhysRevLett.122.246401,PhysRevLett.125.257602,PhysRevB.103.205413}. Symmetry-lowering effects of the full model then determine the ground state on smaller energy scales, while external perturbations (such as disorder, stain, or substrates) can change the selection \cite{PhysRevB.100.125104,PhysRevB.103.125138,PhysRevLett.127.027601,PhysRevX.11.041063,Wagner2022Apr}. 
However, internal processes, such as the mutual feedback between correlations of different ordering channels, can be equally decisive in this situation. The near-degeneracy of ordered states that was demonstrated in Hartree-Fock calculations\cite{
Bultinck2020Aug,PhysRevLett.124.097601,Liu2021Jan,PhysRevLett.127.027601,PhysRevX.11.041063,Wagner2022Apr,kwan2023electronphonon,kwan2023electronphonon} highlights the need for unbiased treatments of competing orders. Recently, Monte Carlo simulations\cite{PhysRevX.11.011014,PhysRevX.12.011061,Pan2022Mar,Zhang2021Jul} and renormalization group calculations\cite{PhysRevResearch.4.033168,Throckmorton2020May} addressed this competition, but within different models for the narrow bands in TBG, which affects the delicate interplay of orders.

In this work, we present a complementary analysis that investigates competing orders as instabilities of itinerant Dirac electrons. Focusing on the Dirac spectrum and employing an unbiased renormalization group approach allows us to draw universal conclusions and assess relevant orders. 
Our rationale comes from the observed signatures of a Dirac dispersion in TBG, and the relative reduction of interaction effects due to the vanishing density of states of Dirac electrons. 

As a point of departure, we use an effective low-energy model valid around charge neutrality that respects the symmetries and topology of the narrow bands of TBG in the vicinity of the magic angle \cite{Maine2020Nov}. We argue that a chiral SU(4) symmetry emerges in this low-energy Dirac fermion model, which we generalise to an arbitrary number of flavours (Dirac points) $N_f$. 
We study possible phase transitions 
based on all symmetry-allowed local interactions 
via the perturbative renormalisation group (RG).
To do so, we analyse the fixed point structure of the RG equations for a Fierz-complete basis of the fermionic interactions.  
Our analysis is valid beyond the concrete application to TBG for other two-dimensional Dirac systems and thereby complements, e.g., previous fermionic RG investigations of phase transitions in graphene\cite{PhysRevLett.97.146401,Semenoff_2012,PhysRevB.79.085116,PhysRevB.103.205135,PhysRevD.82.085018,PhysRevD.92.085046,PhysRevB.107.035151,Roy2019Mar}. 
We classify possible quantum phase transitions based on flavour and Lorentz symmetry of their order parameter manifolds, and if they dynamically generate a mass that gaps out the Dirac spectrum or if they distort the spectrum but maintain (semi-)metallicity. 

We find that, 
although there are many accessible transitions on the single-channel mean-field level, which is equivalent to large flavour numbers, the corresponding fixed points can disappear for small flavour numbers due to the coupling between different ordering channels. As a result, only a few transitions remain accessible for small flavour numbers, which is relevant to TBG. They include transitions with emergent Lorentz symmetry to a quantum anomalous Hall state and 
as well as a state with an SU(4) order parameter manifold that
contains spin Hall, valley (spin) Hall and 
intervalley-coherent states with a gapped spectrum.
We also determine how relaxing SU(4) symmetry affects the possible phase transitions. We show that the instabilities descending from the order parameter manifold of insulating states remain accessible separately. In addition, we argue that perturbing SU(4) stabilizes instabilities towards spin-(valley-)-polarized, spin nematic, and 
metallic intervalley-coherent states.

\section{SU(4) Dirac electrons}
\subsection{Dirac fermion model of twisted bilayer graphene}
\label{sec:DiracTBG}

Our starting point is a low-energy effective model for Dirac fermions around charge neutrality in TBG\cite{Maine2020Nov} with the action at zero temperature 
\begin{equation}\label{h0}
    \mathcal S_0=\int\!\!\! \frac{d^3q}{(2\pi)^3} \,\psi^{
    \dagger}(-i\omega+\rho_x\tau_z q_x +\rho_y q_y)
    \psi
    \,,
\end{equation}
 where $\psi$ is a 16-component spinor due to spin, sublattice, 
 valley, and mini-valley degrees of freedom. We denote unity ($\gamma=0$) and Pauli ($\gamma={x,y,z}$) matrices in sublattice space with $\rho_\gamma$ and in valley space with $\tau_\gamma$. Analogously, we use $\sigma_\gamma$ and $\mu_\gamma$ for spin and mini-valley space below. 
 The sublattice and valley degrees of freedom form the Clifford algebra of the Dirac spinors (see below for a Lorentz-invariant formulation), while the kinetic part of the action $\mathcal S_0$ is diagonal in spin and mini-valley space. 
 
 The mini-valley degree of freedom originates from the Dirac points in the same valley of the two different graphene layers, i.e. the two mini-valleys per valley possess the same chirality. 
  
 We generalize the mini-valley degrees of freedom to an arbitrary number of 
 flavours for the Dirac fermions $\psi\rightarrow\psi^\alpha$ with $\alpha=1,2,\ldots N_f$. In the case of TBG $N_f^{TBG}=2$, i.e., $N_f$ counts the number of eight-component spinors \footnote{Note that this deviates from other common definitions which count two- or four-component spinors.}. 
The Dirac fermion model \eqref{h0} implements the emergent $C_2$ symmetry of the continuum model of TBG via $R_{C_2}=\rho_x \tau_x\mu_x$, $\bm{q} \rightarrow -\bm{q}$.  and time-reversal via $\Theta=\sigma_y \tau_x \mu_x \mathcal K$, $\bm{q} \rightarrow -\bm{q}$. The threefold rotation $C_3$ is promoted to a full U(1) 
rotational symmetry $R_{rot}=\exp[-i \varphi \rho_z\tau_z]$ in the low-energy limit around charge neutrality.  
Furthermore, since the hybridization between layers can be treated as a perturbation for states around the Dirac cones\cite{PhysRevLett.99.256802}, the spin-valley SU(2)$\times$SU(2)$\times$U(1) symmetry of TBG is enhanced to an emergent SU(4) in the low-energy limit, similar to the chiral limit\cite{PhysRevLett.122.106405,PhysRevB.103.205413}. It is generated by the set of matrices 
\begin{equation}\label{eq:SU4}
   T=\{\sigma_c,\tau_z\sigma_{\gamma},\rho_y\tau_x\sigma_{\gamma},\rho_y\tau_y\sigma_{\gamma}\} 
\end{equation}
with $c=\{x,y,z\}$ and $\gamma=\{0,x,y,z\}$. 
The generalization of the flavour index leads to a unitary flavour symmetry acting via a transformation $U_f \in$ U$(N_f)$. 

On the level of interactions, we include all symmetry-allowed local four-fermion couplings according to the symmetries above. They will be generated by fluctuations, even if zero in the microscopic Hamiltonian so that it is important to include them for an unbiased analysis.

In particular, this means that we must include couplings that break the U(4)$\times$U(4) symmetry of the interactions in TBG in the chiral limit because it is 
partially broken
by the dispersion\cite{Bultinck2020Aug,PhysRevLett.122.246401,PhysRevLett.125.257602,PhysRevB.103.205413}. 
We discuss effects that break chiral SU(4) symmetry further in Sec.~\ref{sec:perturbations}. 
We also retain the emergent flavour symmetry in the interactions, i.e., 
different symmetry breaking patterns within mini valleys (which would translate to layer polarization or translational symmetry breaking on the scale of the moir\'e lattice) are degenerate. 
We then find that there are six distinct 
 couplings allowed by these symmetries so that the interaction Lagrangian is given by 
\begin{align}\label{lag}
\mathcal{L}_{int} &= 
  \frac{g_1}{N_f}(\psi^{\alpha\dagger}\psi^{\alpha})^2+\frac{v_1}{N_f}\sum_{i=1}^{15}(\psi^{\alpha\dagger}T_i\psi^{\alpha})^2  \nonumber \\ & +\frac{g_4}{N_f}(\psi^{\alpha\dagger}\rho_z\tau_z\psi^{\alpha})^2 +\frac{v_4}{N_f}\sum_{i=1}^{15}(\psi^{\alpha\dagger}\rho_z\tau_z T_i\psi^{\alpha})^2 \nonumber \\
  & + \frac{g_2}{N_f}(\psi^{\alpha\dagger}\boldsymbol{\nu}\psi^{\alpha})^2 + \frac{v_2}{N_f}\sum_{i=1}^{15}(\psi^{\alpha\dagger}\boldsymbol{\omega}_i\psi^{\alpha})^2\,
\end{align}
where $T_i\in T$,  $\boldsymbol{\nu}=(\rho_x\tau_z,\rho_y)$, and  $\boldsymbol{\omega}_i  =(\rho_x\tau_zT_i,\rho_yT_i)$.

We assume the long-range Coulomb tail to be screened by the surrounding gates. 
The 64 matrices describing the couplings in Eq.~\eqref{lag} form a complete basis set for 8$\times$8 matrices. Furthermore, we only included flavour-diagonal 
terms in $\mathcal L_{int}$ because Fierz identities allow us to appropriately rewrite any flavor-symmetry-allowed terms as a linear combination of the ones considered 
in Eq.~(\ref{lag}) (see App.~\ref{app:Fierz}). Note that this includes terms with explicit mini-valley dependence $(\psi^\dagger M \boldsymbol{\mu} \psi)^2$, where $M\in \{T_i, \rho_z\tau_z,\rho_z\tau_z T_i,\boldsymbol{\nu},\boldsymbol{\omega}_i\}$.
Thus, the six couplings form a Fierz-complete basis for interacting SU(4)$\times$SU($N_f$)-symmetric Dirac fermions with $N_f>1$.
The inclusion of all terms allowed by the Fierz identities is important for the correct identification  of critical points\cite{Braun_2012,Gehring2015Oct}. 
For $N_f=1$, there are additional Fierz identities which reduce the number of couplings to three (see App.~\ref{app:Fierz}).

In the following we study the full effective action 
\begin{equation}\label{totgamma}
    \Gamma=\mathcal S_0+\int \!\!\!d^3\Vec{x}\mathcal{L}_{int}
\end{equation}
We focus on particle-hole instabilities motivated by the experimental observations of semimetallic or insulating ground states at charge neutrality in TBG. Pairing instabilities are, in principle, also contained in our setup and can be analysed via the use of other Fierz identities that translate to the pairing channel.

\subsection{Ordered states}
\label{sec:orders}
Taken on their own, each of the couplings can induce an instability towards a state with spontaneously broken symmetry if strong enough. The ordered states are characterized by the condensation of the corresponding bilinears $\phi_M=\langle\psi^{\alpha\dagger} M \psi^\alpha \rangle\neq 0$ with $M\in\{T_i,\rho_z\tau_z,\rho_z\tau_zT_i,\boldsymbol{\nu},\boldsymbol{\omega}_i\}$ (see also Tab.~\ref{Table:1}). Note that $n=\langle\psi^{\alpha\dagger} \psi^\alpha\rangle$ is fixed by the density and does not break any symmetry. The condensation of $\phi_{\rho_z\tau_z}$ corresponds to the spontaneous formation of a quantum anomalous Hall state (QAH) and generates a gap in the spectrum. A finite $\phi_{\rho_z\tau_zT_i}$ also gaps out the Dirac fermions. 
Its SU(4) order parameter manifold contains quantum spin Hall (QSH) $\sim\rho_z\tau_z\sigma_c$, valley 
Hall (VH), and valley spin Hall (VSH) $\sim\rho_z\sigma_\gamma$ states, as well as inter-valley coherent order \cite{PhysRevX.8.031089} 
$\sim\rho_x\tau_{x,y}\sigma_\gamma$ (IVC-1) 
that anticommutes with the Dirac Hamiltonian. 
In contrast, the SU(4) order parameter manifold described by $\phi_{T_i}$ splits the degeneracy of the Dirac spectrum, resulting in a metallic state (except the splitting becomes on the order of the bandwidth). It contains spin-polarized $\sim\sigma_c$, valley-polarized $\sim\tau_z$, and spin-valley-polarized $\sim\tau_z\sigma_c$ orders, as well as  
inter-valley coherent order 
$\sim\rho_y\tau_{x,y}\sigma_\gamma$
(IVC-2) that commutes with the Dirac Hamiltonian. 
IVC orders can be classified into time-reversal symmetric T-IVC and non-symmetric Kramers K-IVC \cite{Bultinck2020Aug} states. This classification
generally depends on the projection in $\mu$ space.
With their diagonal mini-valley configuration, IVC-1 is a T-IVC, and IVC-2 a K-IVC state. Instabilities towards other IVC states with triplet mini-valley configuration $\mu_{x,y,z}$ can be probed via making use of the Fierz identities (see Sec.~\ref{susceptibilities}).
Finally, condensation of $\phi_{\boldsymbol\nu}$ and $\phi_{\boldsymbol{\omega}_i}$ spontaneously breaks rotation symmetry. This shifts the position of the Dirac points away from the corners of the Brillouin zone and preserves the semi-metallic state. A finite order parameter $\phi_{\boldsymbol\nu}$ couples to the fermions in the same form like a vector potential $\sim (\rho_x\tau_z,\rho_y)$ which leads to the integer quantum Hall (IQH) effect\cite{PhysRevB.80.205319}. The SU(4) order parameter manifold $\phi_{\boldsymbol{\omega}_i}$ contains a spinful variant of this $\sim (\rho_x\tau_z\sigma_c,\rho_y\sigma_c)$ (S-IQH), in addition to nematic $\sim (\rho_{x},\rho_y\tau_z)$ (NEM) and spin-nematic $\sim (\rho_{x}\sigma_c,\rho_y\tau_z\sigma_c)$ (S-NEM) orders, as well as nematic and spin-nematic intervalley coherent states (N-IVC and S-N-IVC). The latter possess matrix order parameters built of two vectors under rotation $(\rho_z\tau_y\sigma_\gamma,\tau_x\sigma_\gamma)$, $(\rho_z\tau_x\sigma_\gamma,\tau_y\sigma_\gamma)$ which are related by valley U(1)$_v$ \cite{christos2023nodal,Samajdar_2021}.

\subsection{Lorentz-invariant SU(4)-symmetric model}
As Dirac fermions are relevant in a variety of systems besides TBG, we reformulate the action in an explicitly Lorentz-invariant form and compare our results in both cases. To this end, we rewrite $\mathcal{L}_0$ as 
\begin{equation}
\mathcal{L}_0=\Bar{\psi}\gamma_{\mu}q_{\mu}\psi
\end{equation}
where $\Bar{\psi}=\psi^{\dagger}\gamma_0$, $\gamma_1=\gamma_0\rho_x\tau_z$, $\gamma_2=\gamma_0\rho_y$. With the requirement that the $\gamma$ matrices satisfy the Euclidean Clifford algebra $\{\gamma_\mu,\gamma_\nu\}=2\delta_{\mu\nu}$, there are four choices of $\gamma_0 \in \{\rho_z\tau_z,\rho_z, \rho_x\tau_x, \rho_x\tau_y\}$. For $\gamma_0=\rho_z\tau_z$, 
the $\gamma$-matrices form a reducible representation of the  two-dimensional Clifford algebra, 
for the other choices, they form reducible, unitarily equivalent four-dimensional representations. 
To impose Lorentz invariance on the interaction Lagrangian \eqref{lag}, we additionally require $g_1=-g_2=:g$ and $v_1=-v_2=:v$.  This yields
\begin{align}\label{lagLorentz}
\mathcal{L}_{int}^{L} &= 
  \frac{g}{N_f}(\Bar\psi^{\alpha}\gamma_\mu\psi^{\alpha})^2+\frac{v}{N_f}\sum_{i=1}^{15}(\Bar\psi^{\alpha}\gamma_\mu T_i\psi^{\alpha})^2  \nonumber \\ & +\frac{g_4}{N_f}(\Bar\psi^{\alpha}\psi^{\alpha})^2 +\frac{v_4}{N_f}\sum_{i=1}^{15}(\Bar\psi^{\alpha}  T_i\psi^{\alpha})^2 
\end{align}
for the 2D representation, or 
\begin{align}\label{lagLorentz2}
\mathcal{L}_{int}^{L} &= 
  \frac{g}{N_f}(\Bar\psi^{\alpha}\gamma_\mu\psi^{\alpha})^2+\frac{v}{N_f}\sum_{i=1}^{15}(\Bar\psi^{\alpha}\gamma_\mu T_i\psi^{\alpha})^2  \nonumber \\ & +\frac{g_4}{N_f}(\Bar\psi^{\alpha}\gamma_{35}\psi^{\alpha})^2 +\frac{v_4}{N_f}\sum_{i=1}^{15}(\Bar\psi^{\alpha} \gamma_{35} T_i\psi^{\alpha})^2 
\end{align}
for the 4D representations, where $\gamma_{35}=i\gamma_3\gamma_5$ is formed by the two additional matrices $\gamma_3,\gamma_5$ which anti-commute with $\gamma_\mu$, i.e. $\gamma_{35}$ commutes with $\gamma_\mu$.
In analogy to the discussion of the ordered states above, the chiral condensates $\langle \Bar\psi^\alpha \psi^\alpha \rangle$ and $\langle \Bar\psi^\alpha T_i \psi^\alpha \rangle$ ($\langle \Bar\psi^\alpha \gamma_{35}\psi^\alpha \rangle$ and $\langle \Bar\psi^\alpha \gamma_{35}T_i \psi^\alpha \rangle$) spontaneously generate a mass in the fermion spectrum, while the vector condensates $\langle \Bar\psi^\alpha \gamma_\mu \psi^\alpha \rangle$ and $\langle \Bar\psi^\alpha \gamma_\mu T_i \psi^\alpha \rangle$ spontaneously break Lorentz symmetry.

\section{Renormalization group analysis}

\subsection{RG flow and beta functions}

To study the possible ordering tendencies, we employ a perturbative renormalization group scheme. 
Within the RG, we successively integrate out modes above a cut-off scale $k$ and express the scale evolution of the couplings $\lambda\in\{g_1,g_2,g_4,v_1,v_2,v_4\}$ via the differential equations $\beta_\lambda=k \partial_k \lambda$. This defines an effective action $\Gamma_k$ at scale $k$, and for $k\rightarrow 0$, we recover the full effective action that includes all quantum corrections. We show in App.~\ref{app:frg} in terms of a functional RG formulation \cite{BERGES2002223,PAWLOWSKI20072831,Kopietzbook,Delamotte2012,Gies2012,Metzner2012Mar,DUPUIS20211} that the one-loop flow equations are independent of the cut-off scheme. For example, a Wilsonian scheme can be used that integrates over modes within a momentum shell. We obtain the flow equations
\begin{align}
\beta_{g_1} &= g_1 - \frac{4}{N_f} [g_1^2 (4 N_f-1)-2g_1g_2-g_1g_4  \nonumber\\
   &-15 g_1(v_1+2 v_2+v_4)-4g_2 g_4-60 v_2 v_4] \label{b1}
\\[10pt]
 \beta_{v_1} &= v_1 + \frac{4}{N_f}  [g_1v_1+2g_2 (v_1+2 v_4)+g_4 v_1+4 g_4 v_2\nonumber\\&
 -4 N_f v_1^2-9v_1^2+30 v_1v_2-
 v_1 v_4+16v_2^2+24 v_2
   v_4-8v_4^2] 
\\[10pt]
  \beta_{g_4}&= g_4 + \frac{4}{N_f} (4 g_1 g_2-3 g_1 g_4- g_2^2+6 g_2 g_4\nonumber\\
 &+12 g_4^2 N_f-3 g_4^2+90 g_4 v_2-45 g_4 v_1-45 g_4 v_4\nonumber\\&
 +60 v_1 v_2-30
   v_2^2) \\[10pt]
    \beta_{v_4}&= v_4+\frac{4}{N_f}  [4 g_1 v_2-3 g_1 v_4+g_2 (4 v_1-4 v_2+6 v_4)\nonumber\\
   &-3 g_4 v_4+12 N_f v_4^2+24 v_1 v_2-13 v_1 v_4\nonumber\\&-12 v_2^2+26 v_2
   v_4+3 v_4^2]\\[10pt]
   \beta_{g_2} &= g_2 -\frac{4}{N_f}  [g_1 (g_2-2 g_4)-4 g_2^2 N_f+g_2 g_4\nonumber\\
   &+15 v_1-15 v_4)+30 v_4 (v_2-v_1)] \\[10pt]
   \beta_{v_2} & = v_2 - \frac{4}{N_f}  [g_1 v_2-2 g_1 v_4+2 g_2 v_4+g_4 (v_2-2 v_1)\nonumber\\
   &-4 N_f v_2^2-8 v_1^2+3 v_1 (5 v_2-4 v_4)-32 v_2^2+13 v_2 v_4-8
   v_4^2] \label{b2}
\end{align}
where we rescaled the couplings $k^{d-2}l_f \lambda \rightarrow \lambda$ with space-time dimension $d=2+1$ and loop integral $l_f$ (see App.~\ref{app:frg}). For Wilson's momentum-shell cut-off, $l_f=2\pi^{d/2}/\Gamma[d/2]$ is the area of the unit sphere in $d=2$ dimensions. 
For the Lorentz-invariant model, we again impose $g=g_1=-g_2$ and $v=v_1=-v_2$. The Lorentz symmetry is maintained along the flow, therefore $\beta_g=\beta_{g_1}=-\beta_{g_2}$ as well as $\beta_v=\beta_{v_1}=-\beta_{v_2}$. 

\subsection{Fixed points and stability}
\begin{figure*}[th]
\centering

    \includegraphics[width=0.45\linewidth]{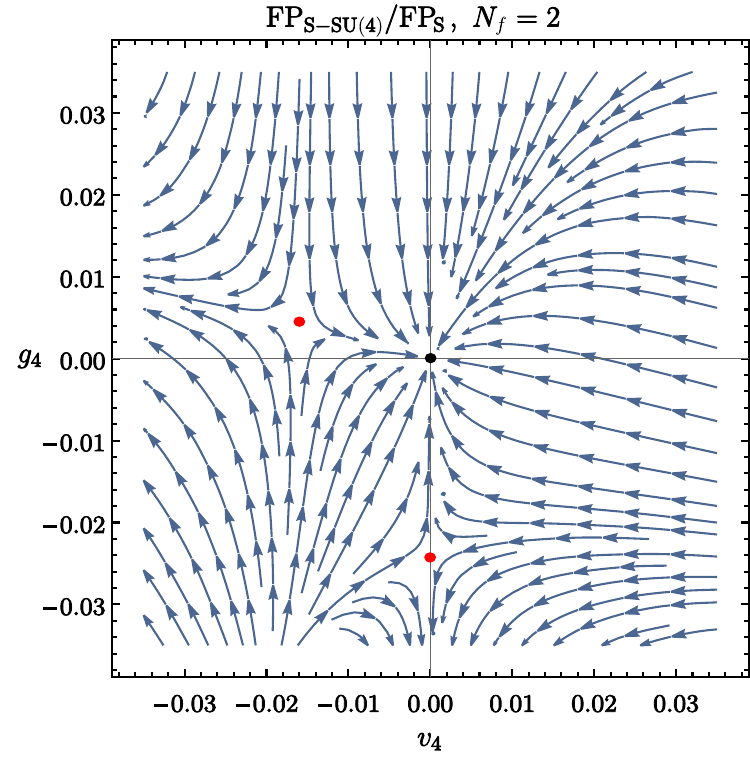}
\quad
    \includegraphics[width=0.45\linewidth]{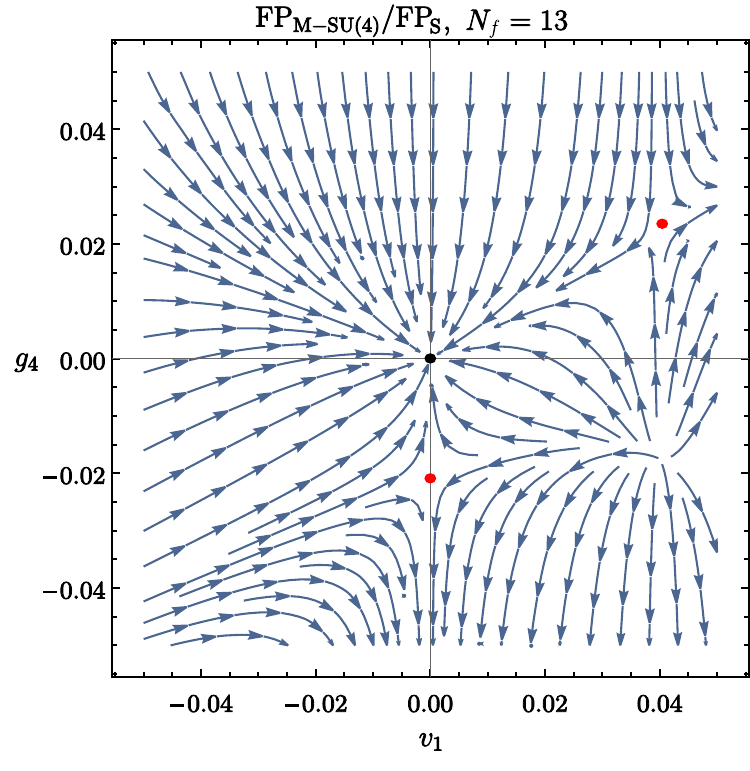}

    \caption{Flow of couplings according to the beta functions $\beta_{g_4},\beta_{v_4}$ and $\beta_{v_1}$ along a plane in coupling space illustrating the behavior around the critical fixed points  $\mathrm{FP}_\mathrm{S}$, $\mathrm{FP}_\mathrm{S-SU(4)}$ and $\mathrm{FP}_\mathrm{M-SU(4)}$ (red)  and the trivial non-interacting Gaussian fixed point (black) for the Lorentz noninvariant case. If the bare values of the couplings are larger in magnitude than the values set by the critical fixed points, they flow to strong coupling in the infrared signaling a potential instability towards symmetry breaking.}
\label{fig:flowdiag}  
\end{figure*}

We are interested in fixed points of the RG equations because they are connected to the possible quantum phase transitions that can be induced by strong couplings. Thus, we look for solutions $\lambda^*=(g_1^*,\ldots,v_4^*)$ where all the beta functions vanish 
\begin{equation}\label{fps}
    \beta_{\lambda_i}(\lambda^*)=0\,.
\end{equation}
To identify which solutions correspond to critical points,
we consider the linearized flow of the beta functions around a fixed point and evaluate the stability matrix
 \begin{equation}
    R_{ij}=-\frac{\partial\beta_{\lambda_i}}{\partial \lambda_j}\Big|_{\lambda=\lambda^*}\,
\end{equation}
which describes how the scale evolution of the couplings is attracted to or repelled from the fixed point. The eigenvalues of the stability matrix determine the critical exponents of the corresponding second-order phase transition. They are universal quantities which do not depend on the microscopic details of the model. 
We are interested in stable fixed points that can be accessed by tuning only one parameter because they are associated to critical points. This means the spectrum of the stability matrix must have all negative eigenvalues except one, which defines a relevant repulsive direction in coupling space.  
This largest critical exponent determines the correlation-length exponent. The second largest exponent describes the corrections to scaling and decides over the stability of the fixed point. 
Fixed points with more than one relevant direction are considered multi-critical or unstable. The trivial Gaussian fixed point ($\lambda=0$), which defines a non-interacting theory in the IR, has a fully negative spectrum, reflecting the need of strong couplings to induce a phase transition. Any non-trivial solution describes an interacting fixed point with finite values of the couplings. If the bare couplings lie beyond the threshold set by the interacting fixed points, their flow diverges along the relevant direction towards the infrared indicating the formation of an ordered state with spontaneously broken symmetry (see Fig. \ref{fig:flowdiag}).

\subsection{Susceptibility analysis}\label{susceptibilities}
\begin{figure*}[th]
\centering
\begin{subfigure}
\centering
    \includegraphics[width=0.49\linewidth]{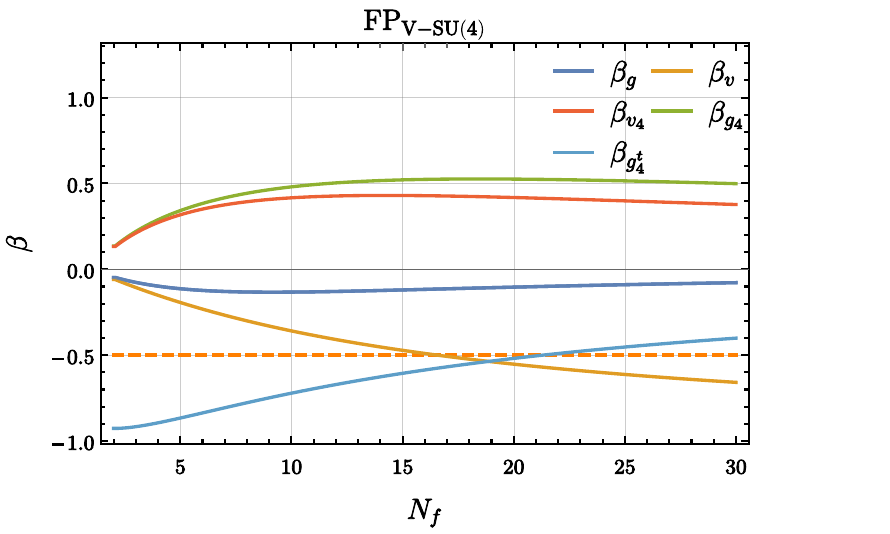}
\end{subfigure}
    \begin{subfigure}
\centering
    \includegraphics[width=0.49\linewidth]{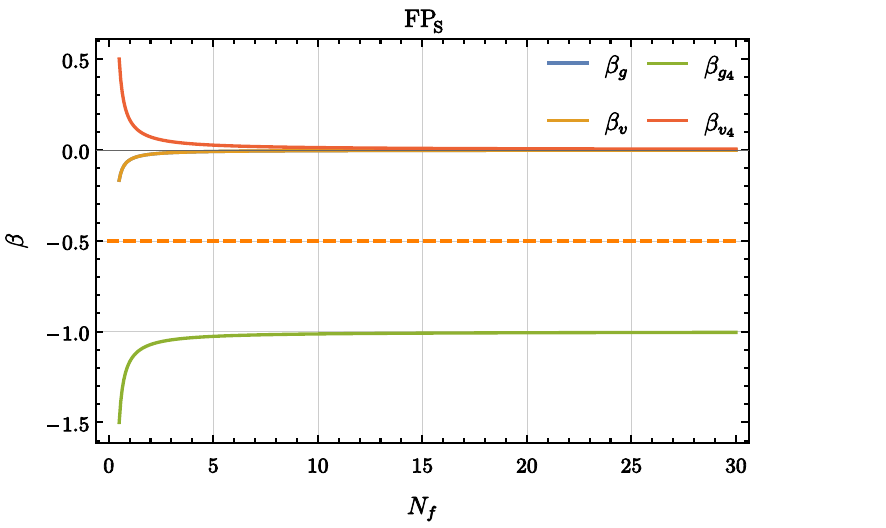}
\end{subfigure}
    \caption{The exponents $\beta$ that determine the divergence of susceptibilities Eq.~\eqref{eq:beta} for 
    the vector SU(4) (FP$_{\mathrm{V-SU(4)}}$) and scalar (FP$_{\mathrm{S}}$) fixed points in the Lorentz invariant case. If $\beta<-\frac{1}{2}$, the corresponding susceptibility diverges signaling a phase transition as marked by the orange dashed line. For the case of $\mathrm{FP}_\mathrm{V-SU(4)}$ the leading instability is in the triplet QAH channel.}

\label{fig:susplotlinv}  
\end{figure*}

The characterization of the possible ordered phases based on the stable fixed points is non-trivial for finite flavour numbers. Since the fixed points' coordinates generally have non-zero values, the identification of the ordering tendencies is complicated by multiple divergent couplings. In order to gain further insight within this fermionic description, we calculate the flow of susceptibilities. This is done by adding an infinitesimal test field 
to the effective action that explicitly breaks the symmetry along a specific ordering channel\cite{PhysRevB.61.7364,PhysRevB.63.035109,PhysRevB.89.201110} 
\begin{equation}\label{sus}
    \Gamma_k \rightarrow \Gamma_k+ h_{i}{\psi^{\alpha\dagger}}{M_i}{\psi^\alpha}\,,
\end{equation}
where $M_i\in \{T_i, \rho_z\tau_z,\rho_z\tau_z T_i,\boldsymbol{\nu},\boldsymbol{\omega}_i\}$. 
These terms represent the coupling of fermions to an external field, which is set to zero at the end.
We define the corresponding susceptibilities as
\begin{equation}\label{sus2}
    \chi_i=
    \frac{\partial^2\Gamma_k}{\partial h_i^2}
    \Bigg|_{h_i=0}\,.
\end{equation}
In the vicinity of the fixed points, the fields and susceptibilities scale according to
\begin{align}
\label{eq:beta}
    & h_i\propto k^{\beta_i}\\
    & \chi_i\propto k^{\gamma_i}
\end{align}
with $\gamma_i=2\beta_i+1$ \cite{PhysRevX.6.041045,PhysRevLett.118.037001,PhysRevB.95.085108}
(see App.~\ref{app:suscept}). 
If $\gamma_i<0$, the corresponding susceptibility diverges. 
In $d=2+1$ dimensions, this means that $\beta<-\frac{1}{2}$. We associate the fixed point with the ordering channel whose susceptibility shows the strongest singularity (Fig.~\ref{fig:susplotlinv}).

We extract the exponents from the flow equations for the infinitesimal fields
\begin{align}
   \partial_t h_{g_1}& =  \frac{2}{N_f} ( g_1 + 2 g_2 + g_4 - 8 g_1 N_f + 15 v_1 \nonumber\\ &+30 v_2 + 15 v_4){h_{g_1}} \label{sus1}\\[10pt]
  \partial_t h_{v_1} &= -\frac{2}{N_f}\left( v_1 + 8 N_f v_1 + 2 v_2 + v_4-g_1 - 2 g_2 - g_4 \right)h_{v_1}
\\[10pt]
 \partial_t h_{g_4}&=  -\frac{6}{N_f}( g_1 - 2 g_2 + g_4 - 8 g_4 N_f + 15 v_1 \nonumber \\& - 30 v_2 + 15 v_4)h_{g_4} \\[10pt]
   \partial_t h_{v_4}&= \frac{6}{N_f} \left( 2 g_2 - g_4 + v_1 - 2 v_2 + v_4 + 8 N_f v_4-g_1 \right)h_{v_4}\\[10pt]
   \partial_t h_{g_2}& = \frac{2}{N_f}\left(g_4 + 8 g_2 N_f - 15 v_1 + 15 v_4-g_1 \right)h_{g_2} \\[10pt]
   \partial_t h_{v_2} &= \frac{2}{N_f}\left( g_4 + v_1 + 8 N_f v_2 - v_4-g_1 \right)h_{v_2}\label{sus6}
\end{align}
at the fixed point solution.
We can see that in the linear response regime the introduction of a symmetry breaking term along a certain channel renormalizes only the respective field with no feedback to the others, i.e the flow equations are decoupled so that we can evaluate them separately. 
Additionally, we note that these equations are Lorentz invariant if we again impose $g=g_1=-g_2$ and $v=v_1=-v_2$.
Analogously, to probe the susceptibilities for flavour-symmetry breaking, we introduce infinitesimal fields $h_i^t\psi^\dagger M_{i} \mu_j \psi$ and calculate the susceptibility exponents. In this case, the flow equations 
are expressed via the triplet couplings $\lambda_i^t$ defined by the interaction terms $(\lambda_i^t/N_f)(\psi^\dagger M_i \boldsymbol{\mu} \psi)^2$ (see App.~\ref{app:suscept}). We obtain the fixed-point values of $\lambda_i^t$ through a Fierz transformation of the fixed points in the flavour-diagonal basis (see App.~\ref{app:Fierz}). 

\subsection{RG flow of the Lorentz invariant system}
\label{sec:Lorentz}

\begin{figure*}[th]
\centering
\begin{subfigure}
\centering
    \includegraphics[width=0.49\linewidth]{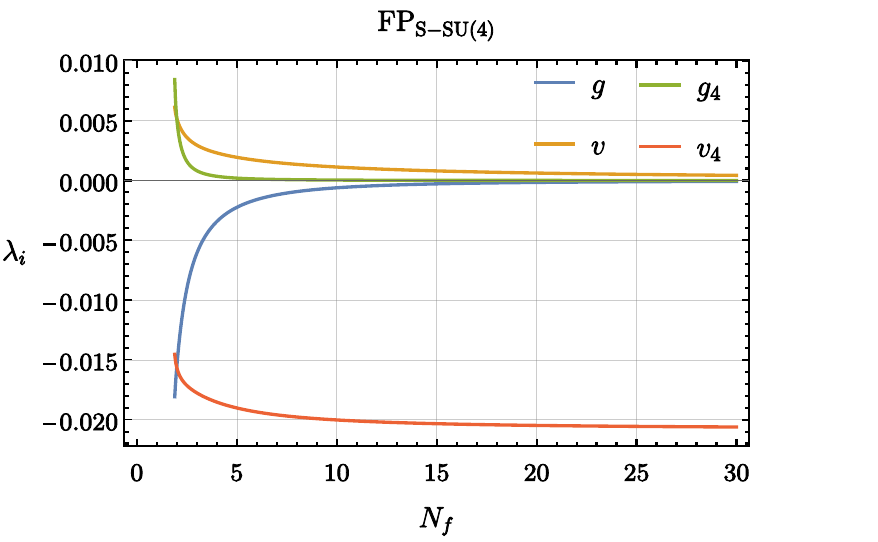}
\end{subfigure}
\begin{subfigure}
\centering
    \includegraphics[width=0.49\linewidth]{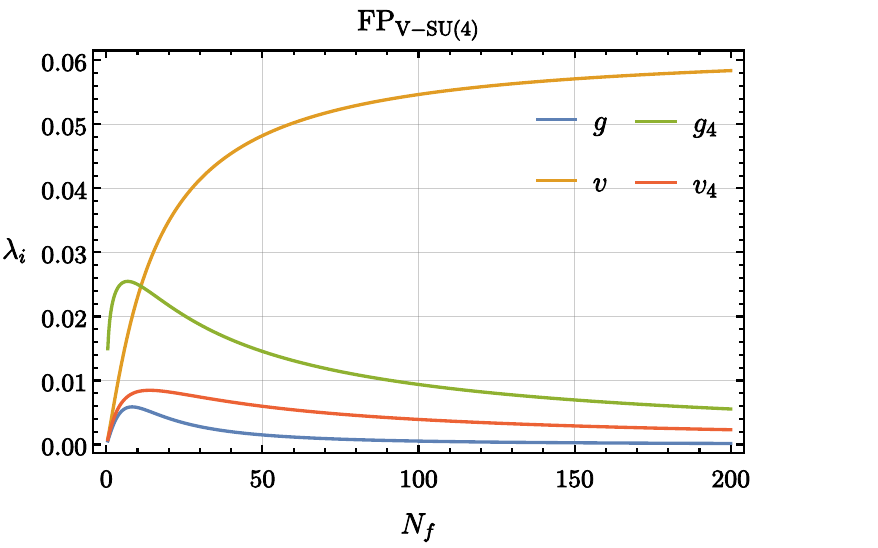}
\end{subfigure}
    \caption{The couplings of the scalar (FP$_{\mathrm{S-SU(4)}}$) and vector (FP$_{\mathrm{V-SU(4)}}$) SU(4) fixed points for several values of the fermion flavor number $N_f$. For general $N_f$, all fixed-point couplings are non-zero demonstrating that different ordering channels are coupled. For large values of $N_f$ only one coupling is non-zero and a single-channel, mean-field description is possible.}
\label{fig:qplor}  
\end{figure*}

\begin{figure*}[th]
\centering
\begin{subfigure}
\centering
    \includegraphics[width=0.49\linewidth]{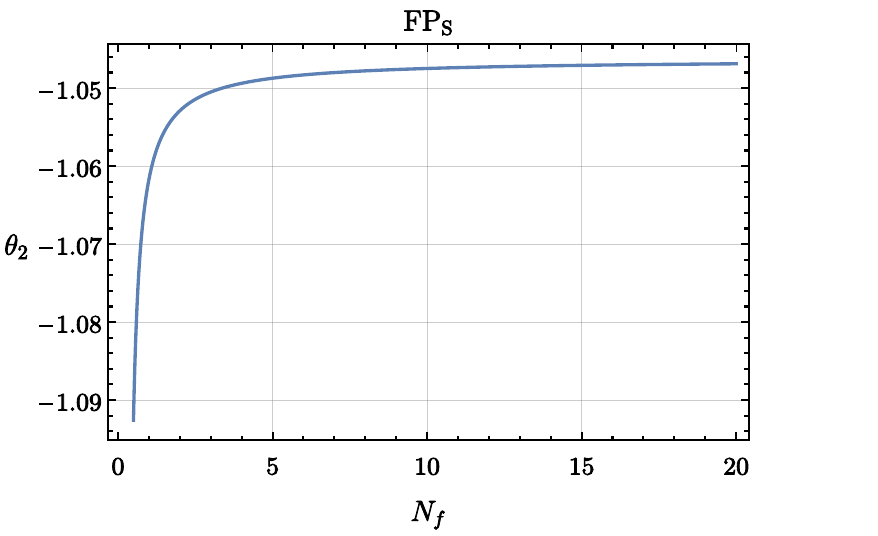}
\end{subfigure}
    \begin{subfigure}
\centering
    \includegraphics[width=0.49\linewidth]{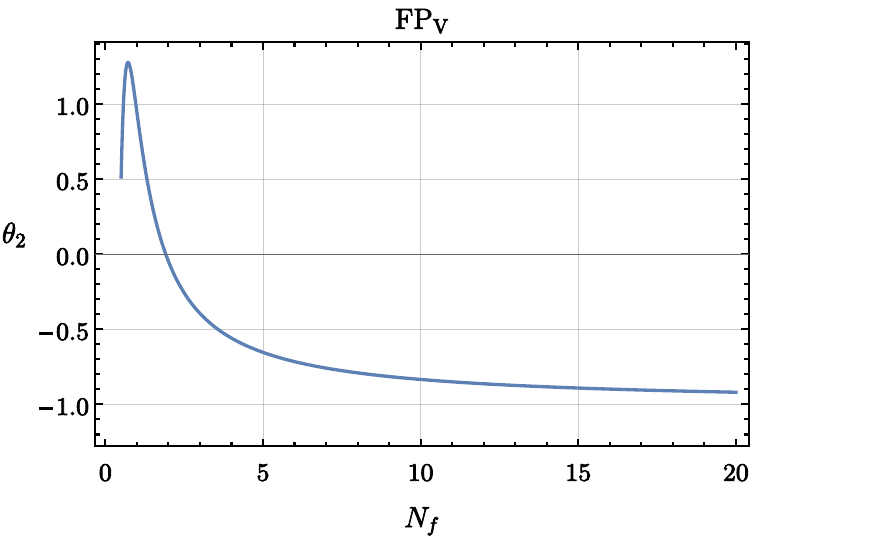}
\end{subfigure}
 \begin{subfigure}
\centering
    \includegraphics[width=0.49\linewidth]{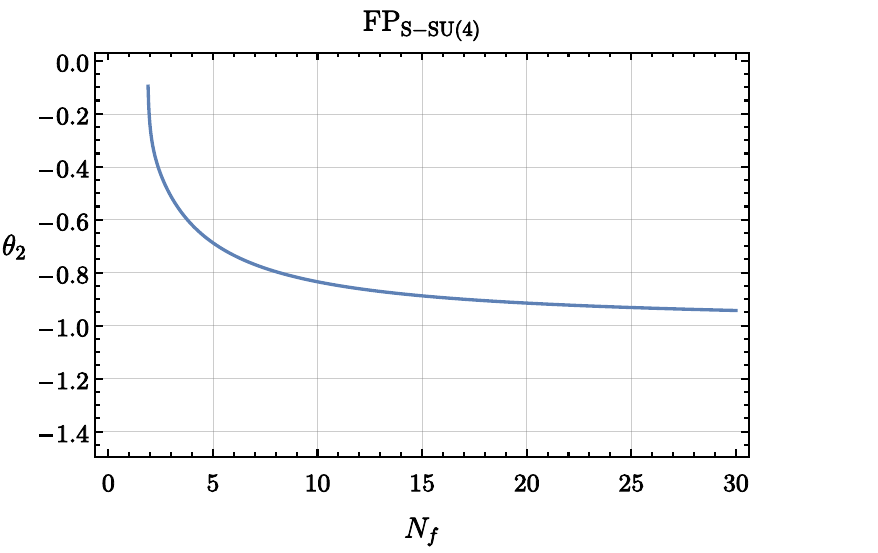}
\end{subfigure}
\begin{subfigure}
\centering
    \includegraphics[width=0.49\linewidth]{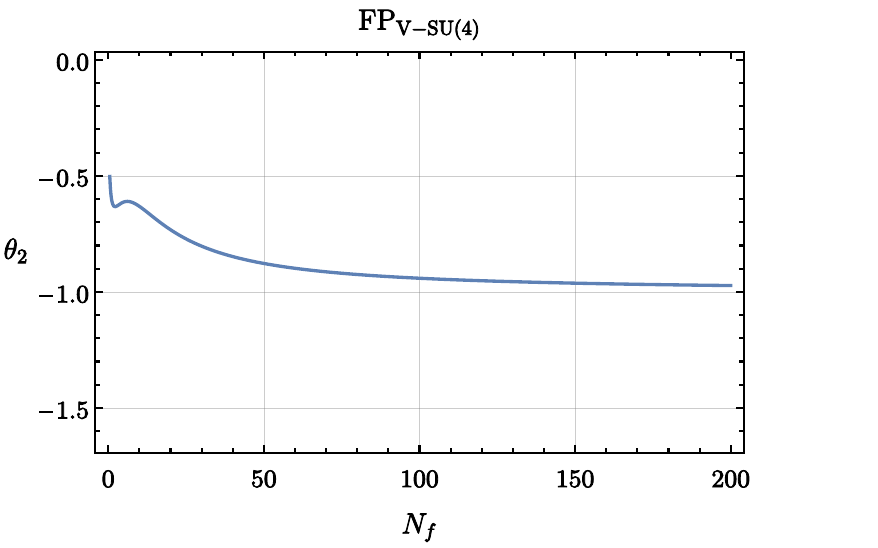}
\end{subfigure}
    \caption{ 
    Second largest critical exponent $\theta_2$ of the scalar (FP$_{\mathrm{S}}$, FP$_{\mathrm{S-SU(4)}}$) and vector (FP$_{\mathrm{V}}$, FP$_{\mathrm{V-SU(4)}}$)  fixed points of the Lorentz-invariant Dirac fermion model. The largest critical exponent $\theta_1=1$ (and is omitted for clarity). The sign of 
    $\theta_2$ dictates the stability of the fixed point solution.  
    It can be clearly seen 
    that $\mathrm{FP}_\mathrm{V}$ becomes unstable at $N_V^c=3$, where $\theta_2$ changes sign.}
\label{fig:qplortheta}  
\end{figure*}

We first analyze the RG flow of the Lorentz invariant system, i.e. we consider the case where $g=g_1=-g_2$ and $v=v_1=-v_2$. Our results for the system without Lorentz invariance and relaxed SU(4) symmetry are presented in the next sections. We perform the fixed point analysis described above for several values of the fermionic flavor number $N_f$. In the large-$N_f$ limit, the RG analysis becomes equivalent to a single-channel, mean-field treatment. But for finite $N_f$, the feedback between ordering channels is important and can qualitatively change possible ordering tendencies as we show below. 
We start with the large-$N_f$ limit, where the beta functions decouple and allow for an unambiguous characterization of the possible ordered phases. 
We identify four stable fixed points that in this limit are characterized by only one coupling of the scalar (S) or vector (V) channels being non-zero 
\begin{align}
    &\mathrm{FP}_{\mathrm{S}}: 
    g_4^*=-\frac{1}{48}+\mathcal{O}(1/N_f)\\
    &\mathrm{FP}_{\mathrm{S-SU(4)}}: v_4^*=-\frac{1}{48}+\mathcal{O}(1/N_f)\\
    &\mathrm{FP}_{\mathrm{V}}: g^*=\frac{1}{16}+\mathcal{O}(1/N_f)\\
    &\mathrm{FP}_{\mathrm{V-SU(4)}}: v^*=\frac{1}{16}+\mathcal{O}(1/N_f)\,.
\end{align}
Upon lowering $N_f$, the coupling between the different ordering channels starts to affect the fixed points and their properties in several ways. With the exception of FP$_\mathrm{S}$, other couplings become non-zero at the fixed points, necessitating the susceptibility analysis for an unambiguous identification of the corresponding instability. 
The scalar fixed point FP$_\mathrm{S}$ is located at 
\begin{equation}\label{eq:FPS}
[g_4^*,v_4^*,g^*,v^*]=\Big[-
\frac{N_f}{12(4N_f-1)},0,0,0\Big]
\end{equation}
for general $N_f$, and the vector fixed point FP$_\mathrm{V}$ at 
\begin{equation}
[g_4^*,v_4^*,g^*,v^*]=[g_{4,V}^*(N_f),0,g_V^*(N_f),0]
\end{equation}
with $g_{4,V}^*(N_f),g_V^*(N_f),0)\geq0$. We give explicit analytical expressions in App.~\ref{app:fps}. For FP$_{\mathrm{S-SU(4)}}$ and FP$_{\mathrm{V-SU(4)}}$, the fixed-point values of all couplings become non-zero for general $N_f$, see Fig.~\ref{fig:qplor}. We find $\lambda_{i,\mathrm{V-SU(4)}}^*\geq 0$ for all $\lambda \in\{g,v,g_4,v_4\}$, while $g_{\mathrm{S-SU(4)}}^*,v_{4\mathrm{S-SU(4)}}^*\geq0$, and $v_{\mathrm{S-SU(4)}}^*,g_{4\mathrm{S-SU(4)}}^*\leq0$. As we described above, the location of the fixed points determines the regime of strong coupling where the flow becomes unstable signalling an instability towards spontaneous symmetry breaking. The flow to strong coupling governed by FP$_\mathrm{S}$ and FP$_{\mathrm{S-SU(4)}}$ is shown in Fig. \ref{fig:flowdiag}. 

Importantly, we observe that not all of the fixed points are accessible for any $N_f$. Specifically, we find critical values of the flavor number $N_f^c$ at which they either disappear or at which they become multi-critical. 
These changes occur via a "collision" with other (multi-critical) fixed points. 
In the first case, the fixed point ceases to exist in the space of real-valued couplings at the critical $N_f$, and instead a pair of complex conjugate solutions moves into the complex plane. 
We find that the fixed-point solution $\mathrm{FP}_\mathrm{S-SU(4)}$ exhibits this behavior and disappears at $N^c_{\mathrm{S-SU(4)}}=1.89$.
In the second case, the second largest eigenvalue of the stability matrix $\theta_2$ changes its sign at the critical value of the flavor number so that the fixed point becomes unstable. The largest eigenvalue of the stability matrix $\theta_1=1$ independent of the critical point and $N_f$ as expected for critical four-fermion models\cite{Gehring2015Oct}. The second largest critical exponent approaches $\theta_2\rightarrow -1$ for large $N_f$, but generally varies as function of $N_f$. 
We show the exponent $\theta_2$  
 for all four critical points in Fig.~\ref{fig:qplortheta}. 
 The quantum critical points labeled as $\mathrm{FP}_\mathrm{S}$ and $\mathrm{FP}_\mathrm{V-SU(4)}$ remain accessible, i.e. $\theta_2<0$, for all values $1<N_f<\infty$, as does $\mathrm{FP}_\mathrm{S-SU(4)}$ in the range where it exists $N_f>1.89$. In contrast, $\mathrm{FP}_\mathrm{V}$ becomes multi-critical at $N^c_\mathrm{V}=3.0$ (see App.~\ref{app:fps} for analytical expression).
 
Furthermore, we underline the importance of including the analysis of the susceptibilities (Sec. \ref{susceptibilities}) in the proper characterization of a fixed-point solution. 

We find divergent susceptibilities at small values of $N_f\approx 2$ for the same instabilities as at large $N_f$ governed by critical points $\mathrm{FP}_\mathrm{S}$, and $\mathrm{FP}_\mathrm{S-SU(4)}$. 
These both correspond to states with a gap in the symmetry-broken regime. In the case of TBG, they describe a QAH 
state, or a state with $\mathrm{SU(4)}$ order parameter manifold connecting 
QSH, VH, VSH, and 
IVC-1 phases. 
In the case of $\mathrm{FP}_\mathrm{V-SU(4)}$, no divergent singlet channel $\sim \mu_0$ exists. Instead we find an instability in the triplet channel, i.e., towards flavour symmetry breaking, $\sim\langle \psi^\dagger \rho_z\tau_z \boldsymbol{\mu} \psi \rangle$, which also opens a gap. In the case of TBG, this describes a QAH state with additional modulation on the moir\'e scale, which includes degenerate layer polarised states $\sim \mu_z$ and moir\'e density waves $\sim \mu_{x,y}$ due to the flavour symmetry.

\subsection{RG flow of the Dirac fermion model for TBG}\label{els}

\begin{figure}
    \centering
    \includegraphics[scale=0.4]{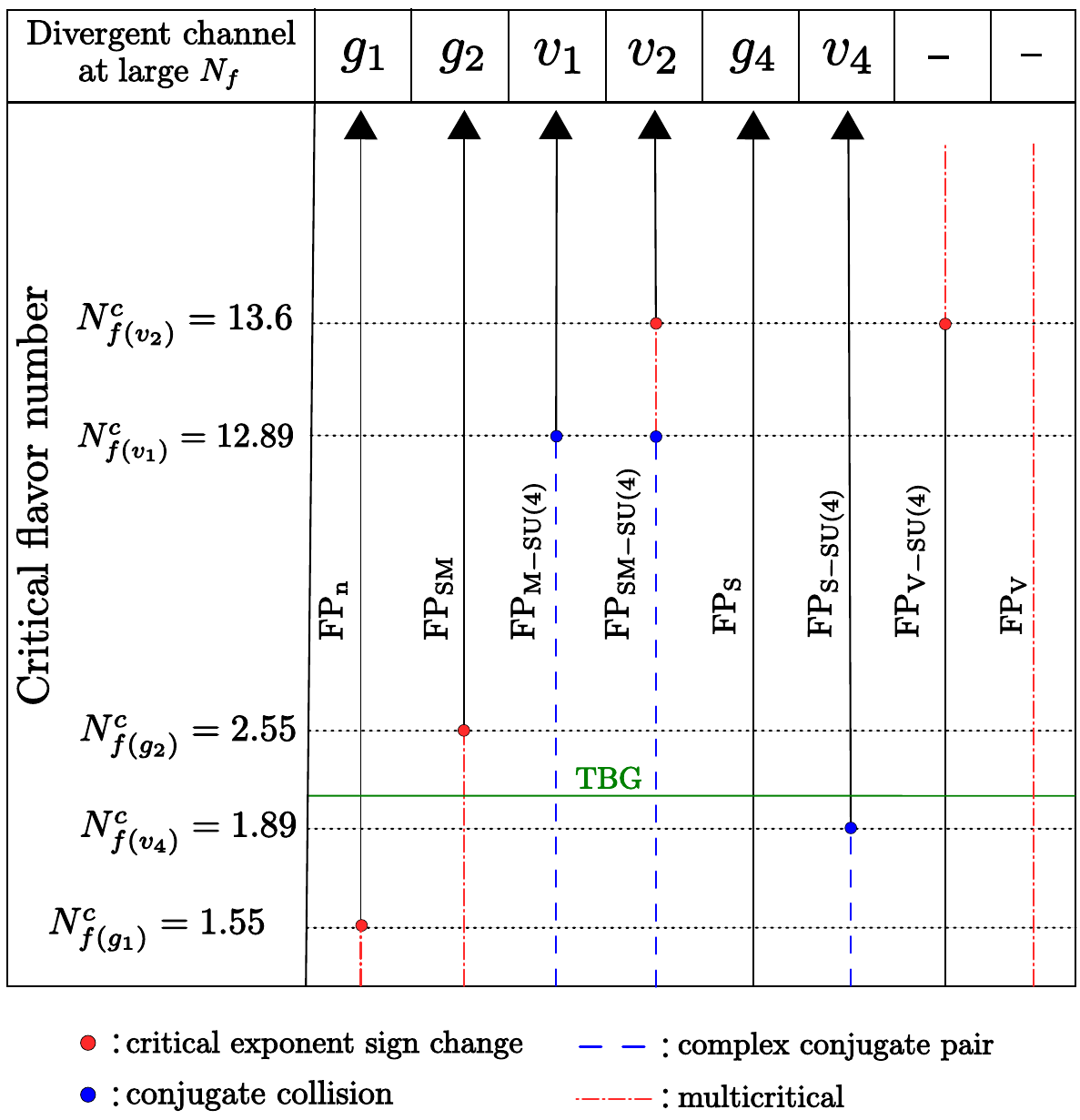}
    \caption{Schematic overview which 
    summarizes the varied behavior of the different fixed point solutions as function of flavor number $N_f$, where $N_f=2$ in the Dirac fermion model of TBG. Density (FP$_n$) and semi-metallic (FP$_{\mathrm{SM}}$) fixed points become multi-critical for small $N_f$, the Lorentz-invariant vector fixed point (FP$_{\mathrm{V}}$) is multi-critical for any $N_f$, and the SU(4) vector fixed point (FP$_{\mathrm{V-SU(4)}}$) at large $N_f$ (red dashed line). Metallic (FP$_{\mathrm{M-SU(4)}}$), semi-metallic (FP$_{\mathrm{SM-SU(4)}}$), and insulating (FP$_{\mathrm{S-SU(4)}}$) SU(4) fixed points lie 
    in the complex plane below a critical flavor number (blue dashed line). Only the scalar fixed point is stable for any $N_f$. It is associated with a QAH phase transition.
    }
    \label{fig:overview}
\end{figure}

Since a physical system like TBG is \emph{a priori} not Lorentz-invariant, we are interested in extending the above discussion to the full Lagrangian spanned by all six couplings in Eq.(\ref{lag}) and compare to what is captured in the Lorentz-invariant case. We thus look to find the fixed point solutions that describe quantum phase transitions for the group of beta functions in Eqs.~(\ref{b1}-\ref{b2}). We find in total seven fixed points that are stable for some range of $N_f$, but generally display a varied behavior as a function of the fermionic flavor number $N_f$. 
In the large-$N_f$ limit, the RG equations again decouple and we can identify six of the seven stable fixed points corresponding to the separate ordering channels
\begin{align}
    &\mathrm{FP}_{\mathrm{S}}: g_4=-\frac{1}{48}+\mathcal{O}(1/N_f) \\
    &\mathrm{FP}_{\mathrm{S-SU(4)}}: v_4=-\frac{1}{48}+\mathcal{O}(1/N_f)\\
    &\mathrm{FP}_{n}: g_1=\frac{1}{16}+\mathcal{O}(1/N_f)\\
    &\mathrm{FP}_{\mathrm{SM}}: g_2=-\frac{1}{16}+\mathcal{O}(1/N_f)\\
    &\mathrm{FP}_{\mathrm{M-SU(4)}}: v_1=\frac{1}{16}+\mathcal{O}(1/N_f)\\
    &\mathrm{FP}_{\mathrm{SM-SU(4)}}: v_2=-\frac{1}{16}+\mathcal{O}(1/N_f) \label{eq:FPSMSU4} \,,
\end{align}
where all other couplings are of order $\mathcal O(1/N_f)$, respectively. We recover the fixed points $\mathrm{FP}_\mathrm{S}$ and $\mathrm{FP}_\mathrm{S-SU(4)}$ of the scalar channels from the Lorentz-invariant case.
The two remaining Lorentz-invariant fixed points
\begin{align}
    &\mathrm{FP}_\mathrm{V}: g_1=-g_2=\frac{1}{16}+\mathcal{O}(1/N_f)\\
    &\mathrm{FP}_\mathrm{V-SU(4)}: v_1=-v_2=\frac{1}{16}+\mathcal{O}(1/N_f)
\end{align}
are unstable for $N_f\rightarrow \infty$. 
Instead, we find the four fixed points $\mathrm{FP}_{n}$, $\mathrm{FP}_{\mathrm{SM}}$, $\mathrm{FP}_{\mathrm{M-SU(4)}}$, and $\mathrm{FP}_{\mathrm{SM-SU(4)}}$ to be stable at large $N_f$. 
FP$_n$ describes the density channel, which does not correspond to any symmetry breaking. A divergence of $g_1$ signals the singular response of the chemical potential to a density change at the Dirac point where the inverse compressibility diverges. 
The semi-metallic (SM) instabilities of FP$_{\mathrm{SM}}$ and FP$_{\mathrm{SM-SU(4)}}$ correspond to the breaking of rotation symmetry with or without an SU(4) order parameter manifold, and FP$_{\mathrm{M-SU(4)}}$ describes polarizing instabilities which yield a metallic (M) state for small order parameters (see Sec.~\ref{sec:orders}).  

In a similar manner to the Lorentz invariant case, the existence and the stability of the fixed points can change when $N_f$ becomes small, and a single-channel mean-field description breaks down see App.~\ref{app:fps} for the evolution of fixed-point couplings and critical exponents with $N_f$). Exceptions are the fixed point related to a QAH instability $\mathrm{FP}_\mathrm{S}$ given by Eq.~\eqref{eq:FPS}, and the fixed point related to the density channel $\mathrm{FP}_n$ given by $g_1^*=N_f/4(4N_f-1)$, which both possess just one non-zero coupling.
A general overview of the of the fixed points' behavior as a function of $N_f$ is provided in Fig.~\ref{fig:overview}. 

We find that $\mathrm{FP}_\mathrm{S}$ and $\mathrm{FP}_n$ remain accessible for any $1<N_f<\infty$. As in the Lorentz-invariant case, $\mathrm{FP}_\mathrm{S-SU(4)}$ exists and is stable for $1.89<N_f<\infty$. This means that Lorentz symmetry emerges in the vicinity of the associated phase transition towards the insulating SU(4) order. In contrast, the Lorentz-invariant fixed point FP$_{\mathrm{V}}$ is always multi-critical if Lorentz invariance is not enforced. 
Interestingly, another fixed point with emergent Lorentz symmetry becomes critical when lowering $N_f$, which is not part of the stable large-$N_f$ fixed points. The Lorentz invariant solution $\mathrm{FP}_\mathrm{V-SU(4)}$ collides with the semi-metallic SU(4) solution $\mathrm{FP}_\mathrm{SM-SU(4)}$
at $N_f^c=13.6$ leaving $\mathrm{FP}_\mathrm{SM-SU(4)}$ unstable and $\mathrm{FP}_\mathrm{V-SU(4)}$ stable at smaller $N_f$. 
At slightly smaller $N_f\approx 12.89$, $\mathrm{FP}_\mathrm{SM-SU(4)}$ and $\mathrm{FP}_\mathrm{M-SU(4)}$ collide and vanish into the complex plane. 
Finally, $\mathrm{FP}_\mathrm{SM}$ becomes multi-critical at $N_f\approx 2.55$. 
We also calculate the exponents characterizing the divergence of susceptibilities for the seven fixed points (see App.\ref{app:suscept}). We find that all the divergent channels can be connected to the large-$N_f$ limit, with the exception of $\mathrm{FP}_\mathrm{V-SU(4)}$, which is not stable at large $N_f$.
Furthermore, slightly before $\mathrm{FP}_\mathrm{SM-SU(4)}$ disappears, the related leading instability changes to the triplet channel (see Fig.~\ref{susplot} in App.~\ref{app:suscept}).

Therefore, a picture similar to the Lorentz-invariant case arises in the Dirac fermion model of TBG at small $N_f\approx 2$. We obtain instabilities related to FP$_\mathrm{S}$  $\mathrm{FP}_\mathrm{V-SU(4)}$ and 
FP$_\mathrm{S-SU(4)}$. These correspond 
to symmetry-broken phases with a gap due to QAH states with ($\sim\mu_0$) or without ($\sim\mu_{x,y,z}$) mini-valley symmetry, or SU(4)-related VH, VSH, QSH and
IVC-1 states. Interestingly, there is neither an instability towards polarizing orders (SP, VP, 
IVC-2) associated with FP$_{\mathrm{M-SU(4)}}$ 
nor towards nematic orders associated with FP$_{\mathrm{SM}}$ or FP$_{\mathrm{SM-SU(4)}}$ in the strict SU(4)-symmetric model.

If however, this symmetry is reduced, 
additional solutions 
become accessible also in the small-$N_f$ limit 
because the critical flavour number of the separate instabilities is lowered. 
This is further elaborated on in the next section. 

\subsection{Relaxing SU(4)}
\label{sec:perturbations}
As described in Sec.~\ref{sec:DiracTBG}, the Dirac fermion model of TBG possesses an emergent SU(4) symmetry, leading to a fourfold degeneracy of the Dirac fermions at charge neutrality. Since three out of the six quantum critical points are related to orders containing an SU(4) symmetric manifold, we are interested in studying whether these fixed-point solutions are robust under a possible breaking of this symmetry. The breaking can occur on the level of the interactions due to high-energy corrections above the UV cutoff of the Dirac fermion model, which lower the symmetry from SU(4) to SU(2)$\times$SU(2)$\times$U(1), i.e., independent spin rotations in both valleys. 

We thus split the initial manifold into five new channels that are invariant by the lowered symmetry 
\begin{align}
    &T^{(1)} =\{\sigma_c\} \nonumber \\
    &T^{(2)}=\{\tau_z\sigma_c\} \nonumber\\ 
    &T^{(3)}=\tau_z \nonumber\\ 
    &T^{(4)}=\{\rho_y\tau_x,\rho_y\tau_y\} \nonumber\\
    &T^{(5)}=\{\rho_y\tau_x\sigma_{c},\rho_y\tau_y\sigma_{c}\} \nonumber
\end{align}
and study the Lagrangian that accounts for the new interaction terms.
The above channel separation means that five new couplings are to be introduced for each term $\propto v_i$ that contained an SU(4) manifold, yielding in total eighteen interaction terms in the lower symmetry Lagrangian
\begin{align}
\mathcal{L}'_{int} &= 
  \frac{g_1}{N_f}(\psi^{\alpha\dagger}\psi^{\alpha})^2+\sum_{i,j}\frac{v_1^{(i)}}{N_f}(\psi^{\alpha\dagger}T^{(i)}_j\psi^{\alpha})^2  \nonumber \\ & 
  +\frac{g_4}{N_f}(\psi^{\alpha\dagger}\rho_z\tau_z\psi^{\alpha})^2 + \sum_{i,j}\frac{v_4^{(i)}}{N_f}(\psi^{\alpha\dagger}\rho_z\tau_z T_j^{(i)}\psi^{\alpha})^2 \nonumber \\
  & + \frac{g_2}{N_f}(\psi^{\alpha\dagger}\boldsymbol{\nu}\psi^{\alpha})^2 + \sum_{i,j}\frac{v_2^{(i)}}{N_f}(\psi^{\alpha\dagger}\boldsymbol{\omega}_j^{(i)}\psi^{\alpha})^2\,
\end{align}
where $\boldsymbol{\omega}_j^{(i)}=(\rho_x\tau_z T_j^{(i)},\rho_y T_j^{(i)})$. We derive the flow equations for the new set of couplings (see App.~\ref{app:su4br}) and investigate how the fixed-point structure changes. 

To study the robustness of the SU(4) symmetry, we evaluate the eigenvalues of the  stability matrix at the coupling coordinates of the SU(4) symmetric quantum critical points.
We find that the fixed points related to $\mathrm{FP}_\mathrm{S}$, $\mathrm{FP}_\mathrm{n}$ and $\mathrm{FP}_\mathrm{SM}$ retain the same behavior as in the previous cases since they are not related to any $\mathrm{SU(4)}$ manifold. 
Furthermore, we consider the stability matrix for $N_f=2, 14, 20$ for the SU(4) fixed points $\mathrm{FP}_\mathrm{M-SU(4)}, \mathrm{FP}_\mathrm{SM-SU(4)}$ and $\mathrm{FP}_\mathrm{S-SU(4)}$, which all become multi-critical. The number of relevant directions is increased to four for $\mathrm{FP}_\mathrm{S-SU(4)}$ and $\mathrm{FP}_\mathrm{SM-SU(4)}$ and to five for $\mathrm{FP}_\mathrm{M-SU(4)}$.  They are primarily along couplings of the channel they descend from, i.e along $v_2^{(i)}$ for $\mathrm{FP}_\mathrm{SM-SU(4)}$, along $v_1^{(i)}$ for $\mathrm{FP}_\mathrm{M-SU(4)}$, and along $v_4^{(i)}$ for $\mathrm{FP}_\mathrm{S-SU(4)}$. 

As such, we observe that the emergent SU(4) symmetry of the Dirac fermion model is not robust against 
 perturbations. Mechanisms which induce a breaking of the $\mathrm{SU(4)}$ symmetry select a specific ground state out of the SU(4) order parameter manifolds. 

To determine possible selections, we follow the $N_f$-evolution of the eighteen single-channel solutions that occur as quantum critical points at large $N_f$ in this case. The result is summarized in Tab.~\ref{Table:1}. 
In total, out of the eighteen large-$N_f$ stable fixed points we find that seven remain critical down to small $N_f\approx2$. Besides $\mathrm{FP}_\mathrm{S}$ and $\mathrm{FP}_\mathrm{n}$, fixed-point solutions characterizing the transition to QSH, VH, VSH, and spinless 
IVC-1 orders are stable. These originate from the $\mathrm{FP}_\mathrm{S-SU(4)}$ fixed point solution in the $\mathrm{SU(4)}$ symmetric case. Lorentz invariance is again emergent at these solutions. 
Furthermore, a fixed point that originates from inaccessible $\mathrm{FP}_\mathrm{M-SU(4)}$ is now stable for $N_f=2$ in the SU(2)$\times$SU(2)$\times$U(1) case. It is related to the transition to a spin-valley polarized state.  
Interestingly, the critical $N_f^c$ for several other fixed points that describe transitions to orders previously related by SU(4) symmetry is also considerably lowered. The effect is the strongest for spin-polarization, 
IVC-2, and N-IVC order. Their critical $N_f$ as part of the SU(4) order parameter manifold $N_f^c\approx 13$ is lowered to $N_f^c\approx 3.66$, $N_f^c\approx 3.2$, and $N_f^c\approx 5.2$, respectively. In this sense, these ordering tendencies are stabilized by perturbations that break SU(4).

\begin{table}[t]
\centering
\caption{
Summary of possible orders introduced in Sec.~\ref{sec:orders} and their tensor structure specified by Pauli matrices in sublattice $\rho$, valley $\tau$ and spin $\sigma$ space ($c=x,y,z$). We also list the corresponding single-channel coupling in the limit of large flavour numbers $N_f$ and the critical 
values $N_f^c$ 
in the $\mathrm{SU(2)}\times\mathrm{SU(2)}\times\mathrm{U(1)}$ case. The dash indicates stability for the whole interval of $N_f$ studied. Double lines group orders related by SU(4) symmetry Eq.~\eqref{eq:SU4}.}
    
\begin{tabular}{ | Sc | Sc | Sc | Sc |} 
\hline 
Order & Tensor structure & Large-$N_f$ 
channel & $N_f^c$  \\
 \hline\hline
  density & $\mathds{1}$ & $g_1$ & 1.55 \\ 
 \hline\hline
  IQH & $\begin{pmatrix}\rho_x\tau_z \\  \rho_y\end{pmatrix}$ & $g_2$ & 2.46 \\ 
 \hline\hline
  QAH & $\rho_z\tau_z$ & $g_4$ & - \\ 
 \hline\hline
  SP & $\sigma_c$ & $v_1^{(1)}$ & 3.66 \\ 
 \hline
 SVP & $\tau_z\sigma_c$ & $v_1^{(2)}$  & 1.48 \\ 
 \hline
 VP & $\tau_z$ & $v_1^{(3)}$  & 4.49 \\ 
 \hline
 
 IVC-2 & $(\rho_y\tau_x,\rho_y\tau_y)$ & $v_1^{(4)}$  & 3.23 \\ 
 \hline
 S-IVC-2 & $(\rho_y\tau_x\sigma_c,\rho_y\tau_y\sigma_c)$ & $v_1^{(5)}$ & 9.14 \\ 
 \hline\hline
 S-IQH & $\begin{pmatrix}\rho_x\tau_z\sigma_c \\ \rho_y\sigma_c\end{pmatrix}$ & $v_2^{(1)}$  & 5.21 \\ 
 \hline
 S-NEM & $\begin{pmatrix} \rho_x\sigma_c \\ \rho_y\tau_z\sigma_c\end{pmatrix}$ & $v_2^{(2)}$ & 2.30 \\ 
 \hline
NEM & $\begin{pmatrix} \rho_x \\\rho_y\tau_z \end{pmatrix}$ & $v_2^{(3)}$ & 6.83 \\ 
\hline
N-IVC & $\begin{pmatrix} \rho_z\tau_y \\ \tau_x \end{pmatrix},  \begin{pmatrix} \rho_z\tau_x \\ \tau_y \end{pmatrix}$ & $v_2^{(4)}$ & 5.23 \\ 
\hline
S-N-IVC & $\begin{pmatrix} \rho_z\tau_y\sigma_c \\ \tau_x\sigma_c \end{pmatrix}, \begin{pmatrix} \rho_z\tau_x\sigma_c \\   \tau_y\sigma_c \end{pmatrix}$ & $v_2^{(5)}$ & 19.9 \\ 
\hline\hline
QSH & $\rho_z\tau_z\sigma_c$ & $v_4^{(1)}$ & - \\ 
\hline
VSH & $\rho_z\sigma_c$ & $v_4^{(2)}$ & - \\ 
\hline
VH & $\rho_z$ & $v_4^{(3)}$ & - \\ 
\hline

IVC-1& $(\rho_x\tau_x,\rho_x\tau_y)$ & $v_4^{(4)}$ & -\\ 
\hline
S-IVC-1 & $(\rho_x\tau_x\sigma_c,\rho_x\tau_y\sigma_c)$ & $v_4^{(5)}$ & 2.75 \\ 
\hline
\end{tabular}
\label{Table:1}
\end{table}

\section{Conclusion}
In the present work we studied competing orders  
in the Dirac fermion model of TBG at charge neutrality in a universal, unbiased way. As such, our considerations generally apply to Dirac fermions with approximate SU(4) symmetry in $2+1$ space-time dimensions. 

We determined a Fierz-complete set of all symmetry-preserving, local 4-Fermi interactions, which we could classify in eighteen, six, or four different channels in the case of SU(2)$\times$SU(2)$\times$U(1), chiral SU(4), or Lorentz invariant SU(4) symmetry, respectively. We calculated the perturbative RG flow of the couplings and investigated their fixed-point structure with the aim of identifying quantum critical solutions that are associated with instabilities towards different ordered states. We diagnosed the instabilities by a divergence in the corresponding susceptibilities. 

The model contains a control-parameter in the form of the fermion flavor number. This allowed us to investigate the interplay of multiple interaction channels, which becomes important in the small-$N_f$ regime relevant to TBG. 
We connected the results to the large-$N_f$ limit, where the  
RG equations decouple and reduce to a single-channel mean-field approach. 

We found a rich landscape of critical fixed points, which display a varied behavior as a function of $N_f$. 
Importantly, we showed that the inter-channel feedback makes many of the single-channel mean-field solutions  
unstable at small $N_f$, either because the solutions disappear or because they become multi-critical. We 
note that, while the critical values of $N_f$ where this happens cannot be determined quantitatively on the one-loop level, the qualitative findings which solutions become unstable are usually robust. 

We  showed that the instabilities that gap out the Dirac spectrum, and thus can gain much condensation energy, 
are particularly stable at  
any $N_f$ and symmetry setups. They correspond to quantum anomalous Hall, 
quantum spin Hall, 
valley (spin) Hall, and time-reversal symmetric inter-valley coherent states in the singlet sector, or a QAH together with moir\'e density waves or mini-valley polarisation in the triplet sector. These solutions possess emergent Lorentz symmetry and thus can already be described via a Lorentz-invariant version of the Dirac fermion model of TBG. The fixed point associated with QAH order remains stable through the entire range of $1<N_f<\infty$, as well as upon breaking SU(4) symmetry. The fixed point related to a 'triplet QAH' instability is only stable below $N_{f}<13.6$ with SU(4) symmetry. Order parameters of QSH, VH, VSH, and IVC-1 form a degenerate manifold in the SU(4)-symmetric case since they are related by symmetry.

The fixed point of the SU(4) manifold $\mathrm{FP}_\mathrm{S-SU(4)}$ disappears at $N_f\approx 1.89$ and we found it to be unstable when SU(4) is broken. 
However, the separate solutions associated with QSH, VSH, VH, and gapped IVC that descend  
from this channel in the SU(2)$\times$SU(2)$\times$U(1) case remain stable in the entire range $1<N_f<\infty$.
The only exception is 
the fixed point that describes a phase transition to a spinful, gapped IVC state, which disappears at $N_f\approx2.75$.

In addition, we found that relaxing SU(4) symmetry stabilizes several orders which generically do not gap out the Dirac spectrum. 
In particular, the critical flavour number of spin-(valley-)polarized, metallic inter-valley coherent, and spin nematic ordering tendencies is strongly reduced. Among them, a spin-valley-polarized state is stable at $N_f=2$. 

Overall, our results provide an unbiased assessment of possible instabilities of multi-flavour Dirac fermions motivated by TBG at charge neutrality. 
In TBG, massless Dirac fermions with reduced velocity are observed for small twist angles \cite{PhysRevLett.117.116804,PhysRevLett.108.076601,PhysRevLett.106.126802,SCHMIDT2010699,PhysRevB.81.165105}. By varying the twist angle, the Fermi velocity changes from zero at the magic angle $\theta\approx 1^\circ$ \cite{doi:10.1073/pnas.1108174108} to the one of graphene when the layers effectively decouple $\theta\gtrsim 20^\circ$ \cite{PhysRevLett.106.126802}. As a result, dimensionless interactions can be tuned from the weakly to the strongly interacting regime via the twist angle. Indeed, strongly correlated behavior was reported for magic-angle TBG \cite{Lu2019,doi:10.1126/science.aaw3780,doi:10.1126/science.aay5533,Stepanov2020,Pierce2021,PhysRevLett.127.197701,Das2021,Wu2021,Polshyn2019,Cao2018Corr,Cao2018Super,doi:10.1126/science.aav1910,Park2021,Zondiner2020,doi:10.1126/science.abb8754,Uri2020,Rozen2021} . Thus, a quantum phase transition of Dirac fermions can be expected at charge neutrality as function of the twist angle.
The instabilities we found that gap out the Dirac spectrum potentially realize the sought-after chiral phase transition in a 2D Dirac material\cite{doi:10.1080/00018732.2014.927109,doi:10.1146/annurev-conmatphys-031113-133841,Boyack2021,Braun_2012,PhysRevLett.97.146401,Semenoff_2012}. The corresponding quantum critical behavior falls into (generalized) Gross-Neveu universality classes\cite{PhysRevD.10.3235,PhysRevB.98.125109,PhysRevB.94.245102,PhysRevD.96.114502,Li_2015,PhysRevD.88.021701,Iliesiu2018,PhysRevB.102.235105,PhysRevD.97.105009,PhysRevD.96.096010,PhysRevLett.123.137602,PhysRevB.97.075129,PhysRevB.96.115132,PhysRevB.97.125137,PhysRevD.103.065018,PhysRevLett.128.225701,Classen_2022}. Interestingly, we also found several critical points whose behavior around a quantum phase transition was not yet studied. 

With respect to selecting the ground state in TBG, we argued that it is crucial to account for the competition between ordering tendencies because several symmetry-broken states lie close in energy. In this context and in light of recent experiments\cite{nuckolls2023quantum,kim2023imaging}, it is interesting to note that  
the leading instability is not necessarily the one expected based on symmetries in the mean-field picture\cite{
Bultinck2020Aug,PhysRevLett.124.097601,Liu2021Jan,PhysRevLett.127.027601,PhysRevX.11.041063,Wagner2022Apr,kwan2023electronphonon} due to inter-channel renormalizations of the interaction. 
It will be enlightening to extend our unbiased treatment by including the breaking of additional symmetries either spontaneously or externally in future studies.
In particular, it was argued that strain plays an important role and selects the so-called incommensurate Kekul\'e spiral (a time-reversal symmetric IVC phase with incommensurate wave vector $q\neq 0$) as the ground state away from charge neutrality \cite{PhysRevX.11.041063,Wagner2022Apr,wang2022kekule,nuckolls2023quantum,kim2023imaging}. Furthermore,
the near-degeneracy of symmetry broken states also suggests an investigation of phases of coexistence.\cite{PhysRevB.84.113404,PhysRevB.90.041413,PhysRevB.92.035429,PhysRevB.93.125119,PhysRevB.102.245128}

\subsection*{Acknowledgments}
We are thankful for valuable discussions with Jens Braun, Pietro M. Bonetti, Steffen Bollmann, Holger Gies, Lukas Janssen, Walter Metzner, Shouryya Ray, Fabian Rennecke, Michael Scherer, and Mathias Scheurer.

\bibliography{bibliography}

\begin{appendix}
\section{Fierz identities}
\label{app:Fierz}
In principle, the generalized flavour symmetry permits terms of the form $\sum_{i=1}^{N_f^2-1}[\Psi^\dagger (M \otimes \kappa_i) \Psi]^2$ in the interacting Lagrangian, where $\Psi=(\psi^1,\ldots,\psi^{N_f})^T$ is a collective $8N_f$-component spinor, $M$ is an $8\times8$ matrix acting in spin-valley-sublattice space, and $\kappa_i$ are generators of SU($N_f$), which replace the Pauli matrices $\mu_c$ of mini-valley space for general $N_f$. To make the flavour index explicit, we can use the completeness relation $\sum_i \kappa_i^{\alpha\beta}\kappa_{i}^{\gamma\delta}=N_f\delta_{\alpha\delta}\delta_{\beta\gamma}-\delta_{\alpha\beta}\delta_{\gamma\delta}$ and rewrite 
\begin{align}
    \sum_{i=1}^{N_f^2-1}[\Psi^\dagger M \otimes \kappa_i \Psi]^2&=-(\psi^{\alpha\dagger} M \psi^\alpha)(\psi^{\beta\dagger}M\psi^\beta)\nonumber\\
    &+N_f(\psi^{\alpha\dagger} M \psi^\beta)(\psi^{\beta\dagger}M\psi^\alpha) \label{triplet_singlet_relation}
\end{align} 
with 8-component spinors $\psi^\alpha$ of flavour $\alpha$. Furthermore, we can connect the 
terms non-diagonal in flavour  $(\psi^{\alpha\dagger} M \psi^\beta)(\psi^{\beta\dagger}M\psi^\alpha)$ to the flavour-diagonal ones 
via Fierz identities using that the 64 matrices $M$ form a basis for $8\times8$ matrices. 
The 
general form of a Fierz identity is 
\begin{equation}\label{eqn:fierz}
    \psi^{\dagger a} M_X\psi^b\psi^{\dagger c}M_X\psi^d = \sum_{Y}F_{XY}\psi^{\dagger a}M_Y\psi^d  \psi^{\dagger c}M_Y\psi^b\,.
\end{equation}
The above equation effectively means that we can write any fermionic bilinear as a linear combination of others provided we know the coefficients $F_{XY}$. To see how diagonal and non-diagonal flavour 
terms are related, we need to set $a=b$ and $c=d$. Thus, the results we provide in the text can be translated to a basis which contains off-diagonal flavour terms.

The Fierz identities can be condensed in the form of a matrix whose entries are exactly the coefficients $F_{XY}$. By labelling the 4-Fermi terms based on the six possible matrix channels $M$, we can  define
\begin{equation}
X_{s/t}=
\begin{bmatrix}
    \boldsymbol{1}_{s/t}\\
    T^i_{s/t} \\
    \rho_z\tau_{zs/t} \\
    \rho_z\tau_zT^i_{s/t} \\
   \nu_{s/t} \\
    \omega_{s/t}   
\end{bmatrix}
\end{equation}
where $M_{s}\equiv(\psi^{\alpha\dagger} M \psi^{\alpha})^2$ and $M_{t}\equiv(\psi^{\alpha\dagger} M \psi^\beta)(\psi^{\beta\dagger} M \psi^\alpha)$,
and Eq. (\ref{eqn:fierz}) can be cast in matrix form  
\begin{equation}
    FX_t=X_s
\end{equation}

with
\begin{equation}
  F=  -\frac{1}{8}
\begin{bmatrix}
1 & 1 & 1 & 1 & 1 & 1 \\
15 & -1 & 15 & -1 & 15 & -1 \\
1 & 1 & 1 & 1 & -1 & -1 \\ 
15 & -1 & 15 & -1 & -15 &1\\
2 & 2 & -2 & -2 & 0 & 0\\
30 & -2 & -30 & 2 & 0 & 0 

\end{bmatrix}\,
\end{equation}

Note that $F^{-1}=F$. Using the completeness relation and the Fierz matrix, we can also rewrite $X_s$ using only triplet terms
\begin{equation}
Y_{t}=
\begin{bmatrix}
    (\psi^\dagger\boldsymbol{1}{\kappa}\psi)^2\\
    (\psi^\dagger T^i{\kappa}\psi)^2 \\
    (\psi^\dagger\rho_z\tau_{z}{\kappa} \psi)^2\\
   (\psi^\dagger \rho_z\tau_zT^i {\kappa}\psi)^2 \\
   (\psi^\dagger\nu{\kappa}\psi)^2 \\
    (\psi^\dagger\omega \kappa\psi)^2
\end{bmatrix}
\end{equation}
via 
\begin{equation}
X_s=\frac{1}{N_f}\left(1-\frac{1}{N_f}F\right)^{-1}FY_t\,.
\end{equation}

In the special case where $N_f=1$, the Fierz matrix relates flavour-diagonal 
terms and the number of independent couplings can be reduced to three, which is the degeneracy of the eigenvalue 1 of the Fierz matrix $F$. 

For the Lorentz-invariant case, where the number of independent couplings is reduced to four, the Fierz identity is given by
\begin{equation}
    F_LX^L_t=X^L_s\,,
\end{equation}

\begin{equation}
X_L=
\begin{bmatrix}
    \gamma_{\mu s/t}\\
    \gamma_{\mu}T^i_{s/t} \\
    \boldsymbol{1}_{s/t} \\
    T^i_{s/t} 
\end{bmatrix}
\end{equation}
and 
\begin{equation}
    F_L= -\frac{1}{8}
\begin{bmatrix}
-1 & -1 & 3 & 3  \\
-15 & 1 & 45 & -3  \\
1 & 1 & 1 & 1  \\ 
15 & -1 & 15 & -1 
\end{bmatrix}
\end{equation}

\section{Derivation of RG equations}
\label{app:frg}
Our starting point is the Wetterich equation\cite{Wetterich1993Feb}
\begin{equation}\label{wett}
    \partial_t\Gamma_k=-\frac{1}{2}\mathrm{Tr}\left\{(\Gamma_k^{(2)}+R_k)^{-1}\partial_t R_k\right\}
\end{equation}
where $\partial_t=k\frac{d}{dk}$ and $\Gamma_k^{(2)}$ is the matrix of functional derivatives of the effective action with respect to the fields, defined as
\begin{equation}
    \Gamma_k^{(2)}={\frac{\Vec{\delta}}{\delta\Psi^{T}(-p)}}\Gamma_k\frac{\lv{\delta}}{\delta\Psi(q)}
\end{equation}
with $\Psi=(\psi,\psi^{\dagger^{T}})^{T}$.
The essence of Eq.(\ref{wett}) is that it describes the evolution of an effective action as a function of a scale variable $k\in [0,\Lambda]$ with $\Lambda$ being a UV-cutoff. This evolution is 
encoded in 
the regulator function $R_t$, which defines the way fluctuations in the interval $[k,\Lambda]$ are integrated out. At $k=0$, all fluctuations are integrated out, which yields the full quantum effective action and the complete solution to the problem.

Based on the discussion of Sec.~\ref{sec:DiracTBG} 
we make an ansatz for the scale dependent action by introducing a scale dependence on all interactions and the regulator function. We neglect the wavefunction renormalization coefficient as diagrams that contribute to the anomalous dimension in a purely fermionic 1-loop approximation vanish in the regime of point-like interactions.
To extract the beta functions for the scale-dependent couplings, we first redefine the scale derivative to only act on the regulator and rewrite Eq.~(\ref{wett}) as
\begin{equation}\label{wett2}
    \partial_t\Gamma_k=-\frac{1}{2}\Tilde{\partial}_t\mathrm{Tr}\left\{\ln{(\Gamma_t^{(2)}+R_k)^{-1}}\right\}\,.
\end{equation}
We split the inverse full propagator into a field-independent part $\Gamma^{(2)}_0$ 
and $\Delta\Gamma_k$ which incorporates the effects of fluctuations. We expand the logarithm around this point
\begin{align}\label{expansion}
    \partial_t\Gamma_k=-\Tilde{\partial}_t\left[
    \frac{1}{2} \mathrm{Tr}\left(\frac{\Delta\Gamma_k}{\Gamma^{(2)}_0}\right)-\frac{1}{4} \mathrm{Tr}\left(\frac{\Delta\Gamma_k}{\Gamma^{(2)}_0}\right)^2 + \ldots\right]\,.
\end{align}
 
We then insert our ansatz and compare 
right and left hand side of Eq.~(\ref{expansion}) 
to identify the beta functions for the running couplings. 

This is analogous to a 1-loop, Wilsonian RG scheme 
for a specific choice of the regulator function. However our analysis will turn out to be independent of the choice of the regulator. Following the procedure detailed above the right hand side of Eq~(\ref{expansion}) evaluates to
\begin{align}\label{betaf1}
   &\partial_t\Gamma_k=  -\Tilde{\partial_t}\int_p\sum_{i,j}2\lambda_i\lambda_j[(\psi^{\dagger} M_iG_kM_j\psi)(\psi^{\dagger}M_jG_kM_i\psi) \nonumber\\
   &+2(\psi^{\dagger}M_i\psi)(\psi^{\dagger}M_jG_kM_iG_kM_j\psi)\nonumber\\
   &-\mathrm{Tr}(G_kM_iM_j)(\psi^{\dagger}M_i\psi)^2 - (\psi^{\dagger}M_iG_kM_j\psi)^2]\,,
\end{align}
where $p$ labels the internal momentum 
integration variable, $M_{i,j}$ are any of the 64 matrices considered in the interacting Lagrangian and $G_k$ is the scale dependent fermionic propagator defined as
\begin{equation}
    G_k=(\Gamma^{(2)}_0+R_k)^{-1}=
    \begin{bmatrix}
        0 & \frac{i\omega +q_x \rho_x\tau_z+q_y\rho_y}{q^2(1+r(k))} \\
        \frac{i\omega +q_x \rho_x\tau_z+\rho_y^Tq_y}{q^2(1+r(k))} & 0 
    \end{bmatrix}
\end{equation}
where $r(k)$ is a dimensionless function encoding the regulator scheme $R_k=(i\omega+q_x\rho_x\tau_z+q_y\rho_y)r(q^2/k^2)$.
Evaluating the matrix products in Eq.~(\ref{betaf1}) yields the beta functions for the six couplings.

The quantity $l_f$ defined as
\begin{equation}\label{lf}
    l_f=-\frac{1}{3}k^{2-d}\Tilde{\partial}_t\int \frac{d^3p}{2\pi^3}\frac{1}{(1+r(p^2/k^2))^2p^2}
\end{equation}
is the threshold function. By rewriting our couplings as $\Bar{\lambda}_i=\lambda_i/l_f$, 
the threshold function is absorbed in the rescaling and thus the beta functions become regulator-independent.

\section{Susceptibilities}
\label{app:suscept}
As mentioned in the main text, to gain insight into the ordered phases related to the quantum critical points, we analyze the divergence of the test-vertex susceptibilities. We start from an effective action
\begin{equation}
    \Gamma_k \rightarrow \Gamma_k+ rh_{i}{\psi^{\alpha\dagger}}{M_i}{\psi^\alpha}+\chi_ih_i^2\,.
\end{equation}
The quantities $r$ and $\chi$ will flow according to:
\begin{align}
    & \partial_tr_i=\frac{\partial}{\partial h_i}\frac{\Vec{\delta}}{\delta \psi^{\mu\dagger}}\partial_t\Gamma_k\frac{\lv{\delta}}{\delta\psi^{\nu}}\Bigg|_{h_i=0,\psi^{\mu\dagger}=0,\psi^{\nu}=0} \label{sus_flows} \\
    &\partial_t\chi_i=\frac{\partial^2(\partial_t\Gamma_k)}{\partial h_i^2}\Bigg|_{h_i=0,\psi^{\mu\dagger}=0,\psi^{\nu}=0}\,,\label{eq:gensus}
\end{align}
where the flow of $\Gamma_k$ is given by Eq.~(\ref{wett}). We perform an expansion of the effective action, while noting that the field-independent part now gets a contribution from the inclusion of the linear vertex term. Keeping only terms that contribute to the flow of the above quantities we get
\begin{align}
    & \partial_tr_i= - l_fC_{ij}\lambda_jr_i\\
    &\partial_t\chi_i=\frac{l_f}{4}r_i^2 \label{chi1}\,,
\end{align}
where $C_{ij}$ is a matrix containing all constant prefactors for each term that appears in the beta functions and $l_f$ is given by Eq.~(\ref{lagLorentz}). Eq.~\ref{sus_flows} is explicitly given by:
\begin{align}
    \partial_t r_i&=-\tilde\partial_t\frac{2\lambda_j}{\dim{M_i}}r_\ell\mathrm{Tr}[M_iM_jG_0M_\ell G_0M_j] \nonumber \\
    &-\mathrm{Tr}[M_iM_j]\mathrm{Tr}[M_jG_0M_\ell G_0] \label{explicit_sus}
\end{align}
We rescale the couplings $\Bar{\lambda}=k^{d-2}\lambda l_f$, with $d$ being the space-time dimension of the model.
At the fixed point solutions, $\Bar{\lambda}_i=\Bar{\lambda}_i^*$, we can solve the differential equations to get the explicit dependence as a function of $k$:
\begin{equation}
    r_i=r_0\left(\frac{k}{\Lambda}\right)^{C_{ij}\Bar{\lambda}_j^*}
\end{equation}
We define $\beta_i\equiv C_{ij}\Bar{\lambda}_i$.  
Then, the dependence of the susceptibility 
can be written as
\begin{equation}
    \chi_i=\chi_0\frac{k^{2\beta+d-2}}{\Lambda^{2\beta}}\,.
\end{equation}
We thus relate the two exponents via $\gamma_i=2\beta_i+d-2$. Setting $d=3$, we get the condition for a divergent susceptibility used in the text $\beta_i<-\frac{1}{2}.$ 
To determine the tendency for breaking of flavour symmetry, we repeat this analysis in the triplet basis, i.e., we introduce test vertices $r_i^th_i^t\psi^\dagger M_i\otimes\kappa_j\psi$ and use the Fierz identities to express the fixed point couplings via combinations of triplet couplings $\frac{\lambda_i^t}{N_f}(\psi^\dagger M_i\otimes \boldsymbol{\kappa}\psi)^2$. 
Replacing $M_i\rightarrow M_i\otimes \kappa_j$ in Eq.~\eqref{eq:gensus}, we find that $\partial_tr_i^t$ can be obtained from $\partial_tr_i$ by replacing singlet via triplet couplings $\lambda_i\rightarrow \lambda_i^t$ and a relative sign between the two terms on the right hand side in Eq.~\eqref{eq:gensus}. Explicitly we obtain
\begin{align}
   \partial_t h_{g^t_1}& =  -\frac{2}{N_f} ( g^t_1 + 2 g^t_2 + g^t_4 + 8 g^t_1 N_f + 15 v^t_1 \nonumber\\ &+30 v^t_2 + 15 v^t_4){h_{g^t_1}} \label{sus1}\\[10pt]
  \partial_t h_{v^t_1} &= \frac{2}{N_f}\left( v^t_1 - 8 N_f v^t_1 + 2 v^t_2 + v^t_4-g^t_1 - 2 g^t_2 - g^t_4 \right)h_{v^t_1}
\\[10pt]
 \partial_t h_{g^t_4}&=  \frac{6}{N_f}( g^t_1 - 2 g^t_2 + g^t_4 + 8 g^t_4 N_f + 15 v^t_1 \nonumber \\& - 30 v^t_2 + 15 v^t_4)h_{g^t_4} \\[10pt]
   \partial_t h_{v^t_4}&= -\frac{6}{N_f} \left( 2 g^t_2 - g^t_4 + v^t_1 - 2 v^t_2 + v^t_4 - 8 N_f v^t_4-g^t_1 \right)h_{v^t_4}\\[10pt]
   \partial_t h_{g^t_2}& = -\frac{2}{N_f}\left(g^t_4 - 8 g^t_2 N_f - 15 v^t_1 + 15 v^t_4-g^t_1 \right)h_{g^t_2} \\[10pt]
   \partial_t h_{v^t_2} &= -\frac{2}{N_f}\left( g^t_4 + v^t_1 + 8 N_f v^t_2 - v^t_4-g^t_1 \right)h_{v^t_2}\label{sus6}
\end{align}

The treatment described above is condensed in Fig. \ref{fig:susplotlinv} in the main text and Figs. \ref{fig:susplotlinv2} and \ref{susplot} in this appendix for the Lorentz invariant model (Eq.~\ref{lagLorentz}) and full SU(4) model (Eq.~\ref{lag}) respectively. We reiterate the importance of this analysis to identify which of the stable fixed points can describe a quantum phase transition. Specifically, as can be seen in Fig.~\ref{susplot} none of the singlet susceptibilities of $\mathrm{FP}_\mathrm{V-SU(4)}$ diverges in the regime of $N_f$ where it is stable. However, 
we identify a divergent susceptibility for $\mathrm{FP}_\mathrm{V-SU(4)}$ that is associated with a triplet QAH state $\langle\psi^\dagger \rho_z\tau_z \boldsymbol{\mu}\psi\rangle$, i.e. a QAH state with additional modulation ($\mu_{x,y}$) or polarisation ($\mu_z$) on the moir\'e scale (see Fig.~\ref{susplot}).
\begin{widetext}

\begin{figure*}[th!]
\centering
 \begin{subfigure}
\centering
    \includegraphics[width=0.49\linewidth]{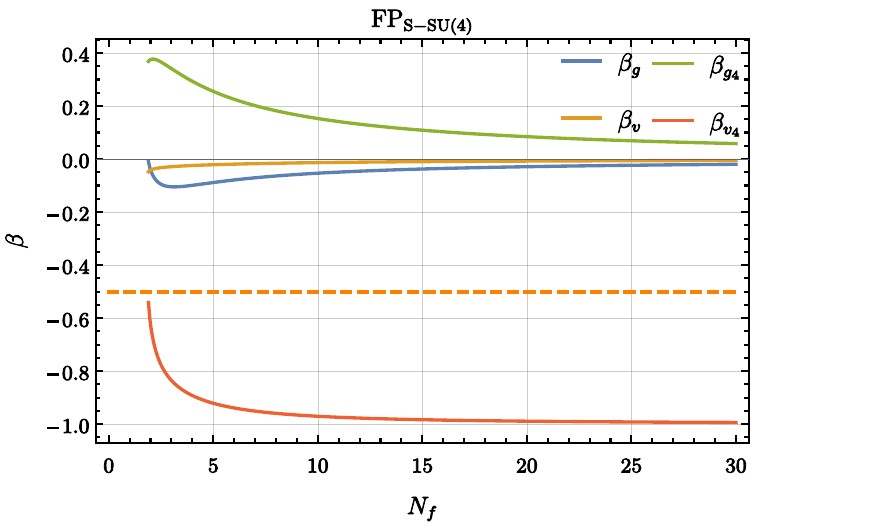}
\end{subfigure}
\begin{subfigure}
\centering
    \includegraphics[width=0.49\linewidth]{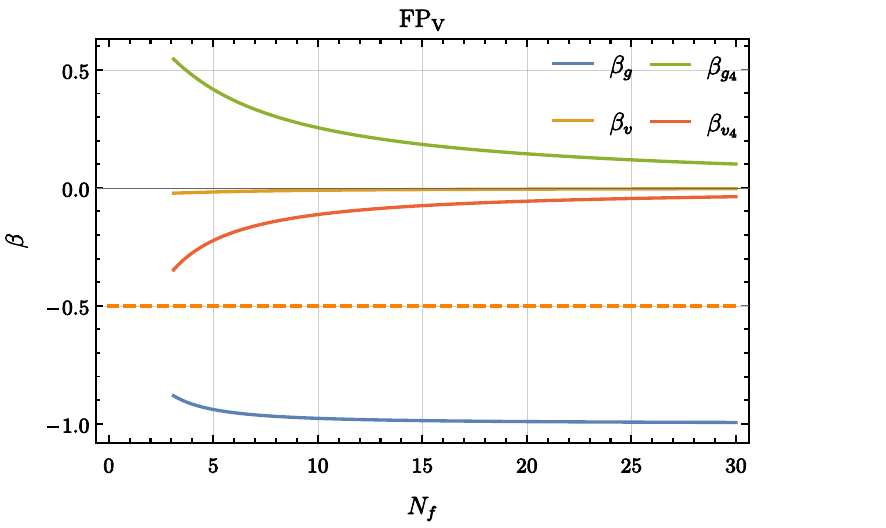}
\end{subfigure}
    \caption{The exponents $\beta$ 
    determining the susceptibilities of the scalar SU(4) FP$_{\mathrm{S-SU(4)}}$ and vector FP$_{\mathrm{V}}$ fixed points in the Lorentz invariant case as function of flavor number $N_f$. If $\beta<-1/2$ the susceptibility of the corresponding order diverges. The exponents for the two other critical fixed points FP$_{\mathrm{S}}$ and FP$_{\mathrm{V-SU(4)}}$ is shown in the main text.}
    \label{fig:susplotlinv2}
\end{figure*}
\begin{figure*}[th!]
\centering
\begin{subfigure}
\centering
    \includegraphics[width=0.49\linewidth]{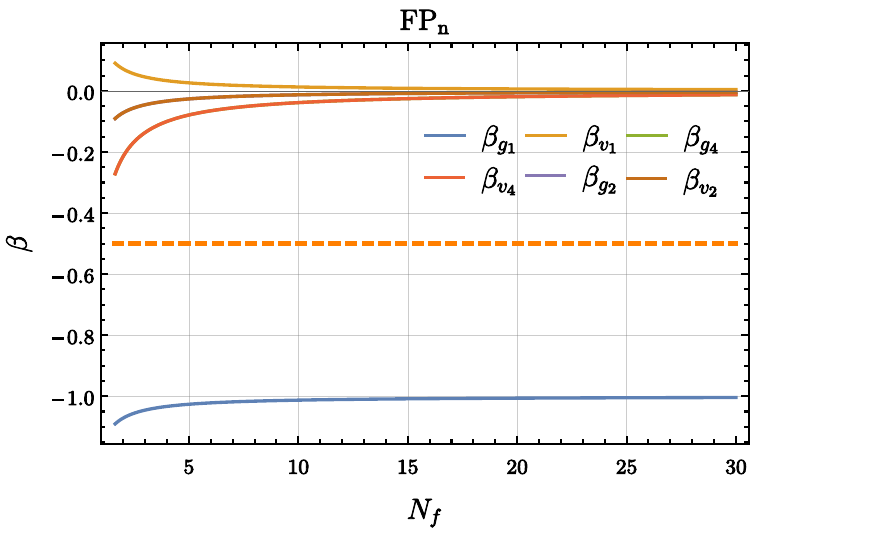}
\end{subfigure}
    \begin{subfigure}
\centering
    \includegraphics[width=0.49\linewidth]{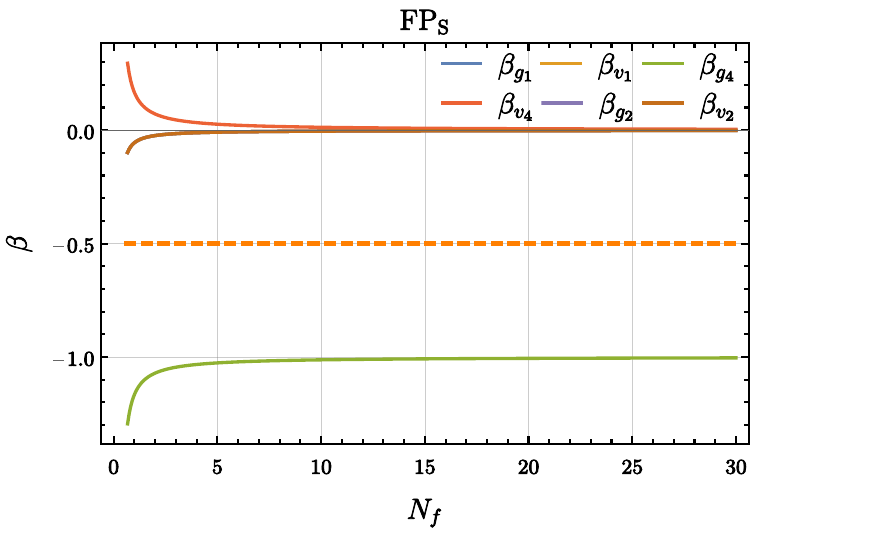}
\end{subfigure}
 \begin{subfigure}
\centering
    \includegraphics[width=0.49\linewidth]{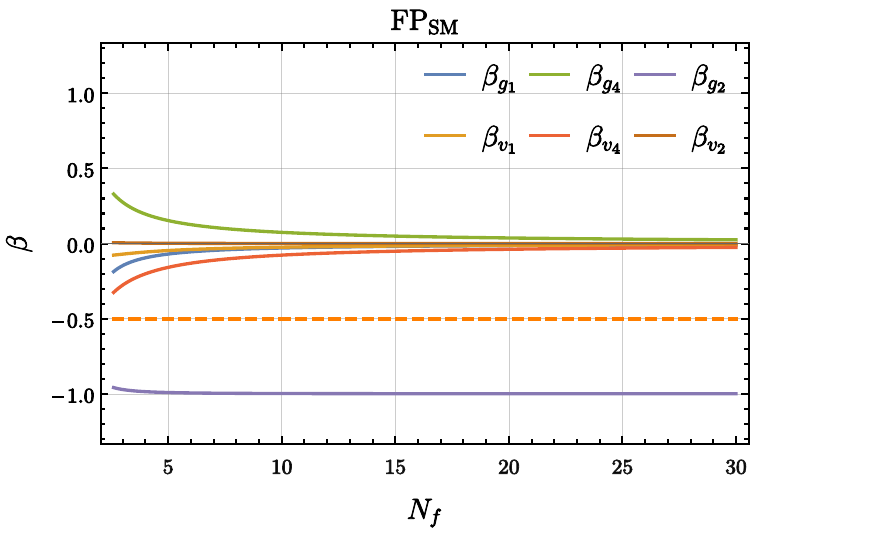}
\end{subfigure}
\begin{subfigure}
\centering
    \includegraphics[width=0.49\linewidth]{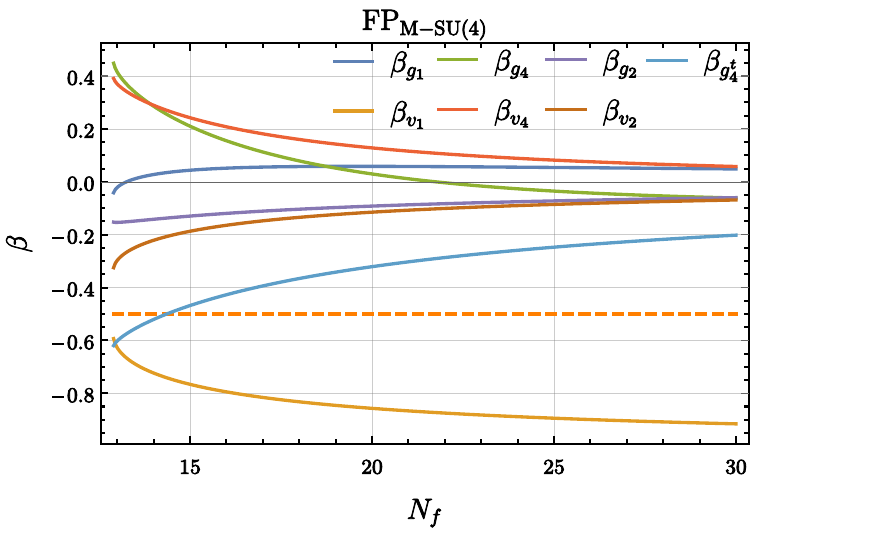}
\end{subfigure}
\begin{subfigure}
\centering
    \includegraphics[width=0.49\linewidth]{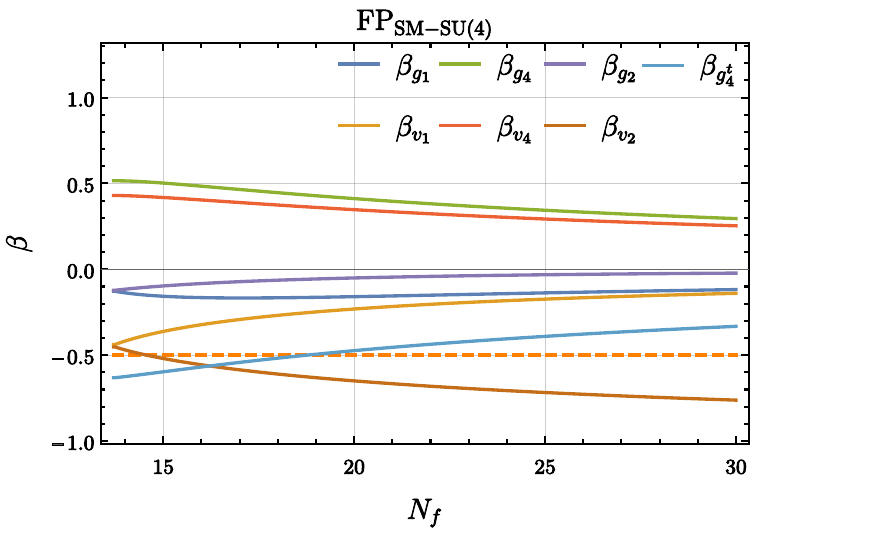}
\end{subfigure}
\begin{subfigure}
\centering
    \includegraphics[width=0.49\linewidth]{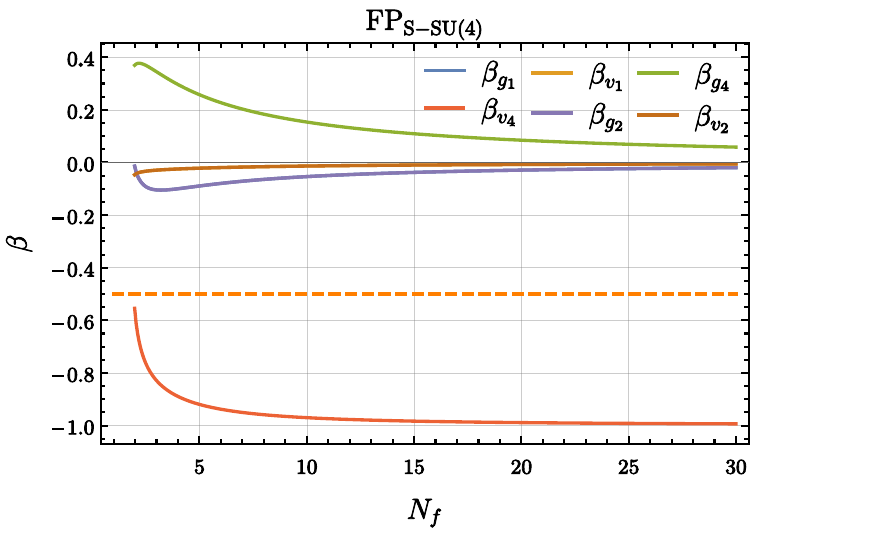}
\end{subfigure}
    \caption{The exponents $\beta$ 
    that determine the behavior of the susceptibilities for all critical fixed points of the SU(4) symmetric Dirac fermion model of TBG as function of flavor number $N_f$. Values of $\beta<-\frac{1}{2}$ can lead to second order phase transitions. All of the fixed point solutions exhibit a leading divergence in a singlet channel with the exception of $\mathrm{FP}_\mathrm{V-SU(4)}$, that has a divergent susceptibility in the triplet QAH channel.}
\label{susplot}  
\end{figure*}
\newpage
\section{Fixed point analysis}
\label{app:fps}
Generally, due to the large number of couplings considered, the fixed point solutions do not allow for an analytic form as a function of $N_f$. However in the case where one or more couplings are zero, the expressions simplify considerably permitting analytic expressions for the quantum critical points. Specifically this holds for $\mathrm{FP}_\mathrm{S}$, $\mathrm{FP}_\mathrm{n}$ and $\mathrm{FP}_\mathrm{V}$, with the respective expressions

\begin{align}
    &\mathrm{FP}_\mathrm{S}: [g_4^*,v_4^*,g^*,v^*]=\Big[-\frac{N_f}{12(4N_f-1)},0,0,0\Big]\\
    & \mathrm{FP}_\mathrm{n}: [g_4^*,v_4^*,g_1^*,g_2^*,v_1^*,v_2^*]=\Big[0,0,\frac{N_f}{4(4N_f - 1)},0,0,0\Big] \\
    & \mathrm{FP}_\mathrm{V}: [g_4^*,v_4^*,g^*,v^*]=\Big[\frac{N_f (-7 + 7 N_f - 
   4 N_f^2 + (1 + 4 N_f) \sqrt{4 + N_f (14 + N_f)})}{4 (-1 + 4 N_f) (5 + 
   4 N_f + 8 N_f^2)},0,\frac{3 N_f^2}{ 4 + 2N_f + 16 N_f^2 + 2\sqrt{4 + N_f (14 + N_f)}},0\Big]   
\end{align}

Fig.~\ref{fig:BIFURC} shows the disappearance of physical solutions into the complex plane related to $\mathrm{FP}_\mathrm{S-SU(4)}$, which constitutes one of the two possibilities that renders solutions inaccessible.

Fig.~\ref{fig:qplor2} shows the behavior of the other two stable fixed points of the Lorentz invariant model and  Fig.~\ref{fig:nonlplots} shows the evolution of the couplings for the fixed points in the full SU(4) symmetric model. These complement Fig.~\ref{fig:qplor} of the main text. Importantly, in Fig.~\ref{fig:nonlplots}, we can observe the recovery of the Lorentz invariant solution $\mathrm{FP}_\mathrm{S-SU(4)}$ in the non-Lorentz invariant case. Furthermore, $\mathrm{FP}_\mathrm{V-SU(4)}$, while accessible for all values of $N_f$ in the Lorentz invariant case, becomes multicritcal in the case of the full SU(4) model (Eq.~\ref{lag}) at $N_f^c\approx 13.89$ as mentioned in the main text. As such, it does not describe a phase transition in the interval of $N_f$ where it is stable.  

\begin{figure*}[h!]
\centering
    \includegraphics[scale=0.8]{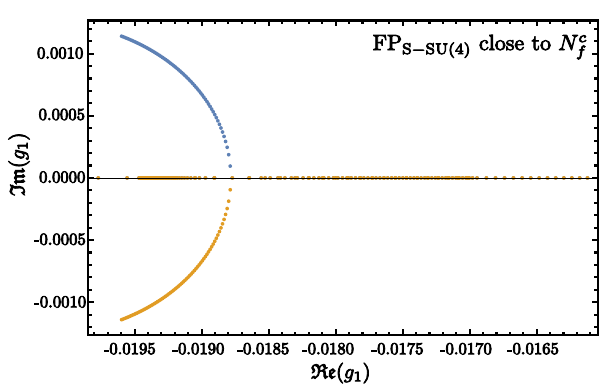}

    \caption{An example of a complex conjugate collision of the scalar SU(4) fixed point $\mathrm{FP}_\mathrm{S-SU(4)}$ with another multicritical fixed point leading to the creation of a pair of real valued fixed point solutions. 
    The points correspond to the coupling values for different $N_f$.}
\label{fig:BIFURC}  
\end{figure*}

\begin{figure*}[h]
\centering
\begin{subfigure}
\centering
\includegraphics[width=0.49\linewidth]{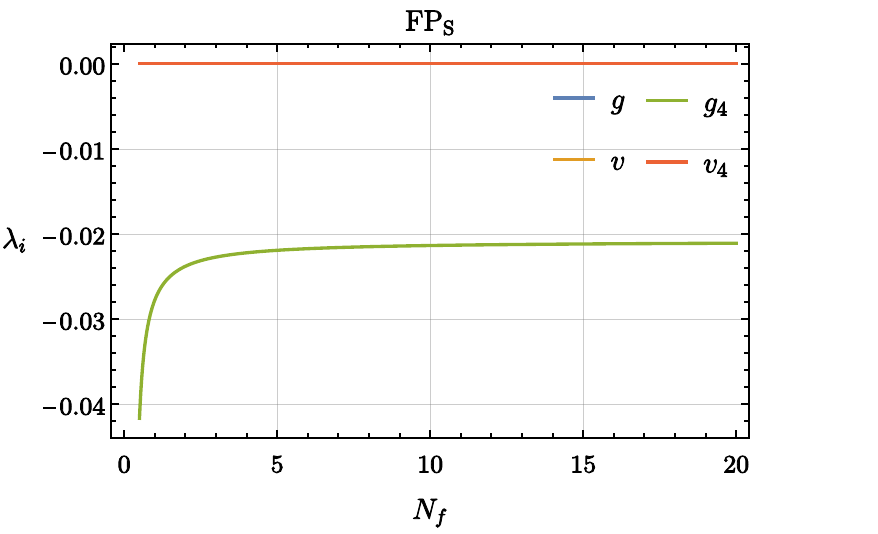}
\end{subfigure}
    \begin{subfigure}
\centering
    \includegraphics[width=0.49\linewidth]{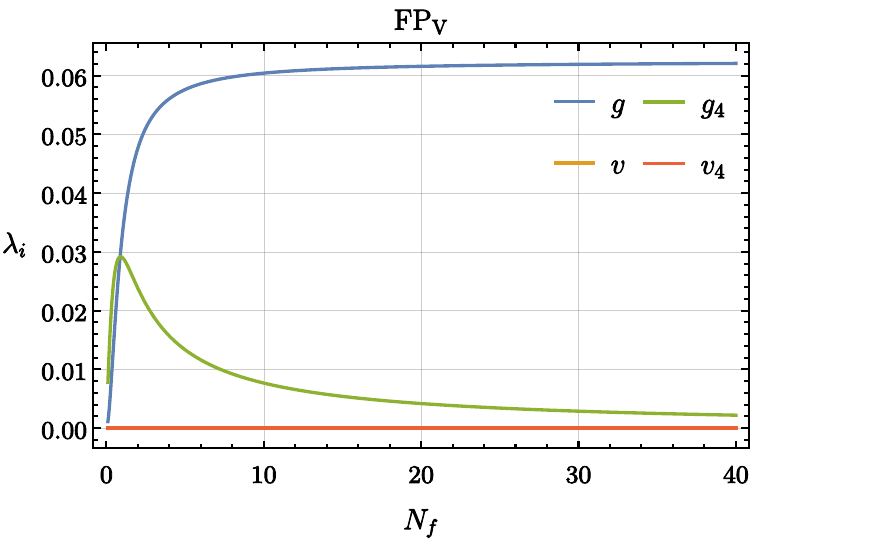}
\end{subfigure}
 \begin{subfigure}
\centering
\end{subfigure}
    \caption{The couplings of the scalar (FP$_{\mathrm{S}}$) and vector (FP$_{\mathrm{V}}$) SU(4) fixed points 
    for several values of the fermion flavor number $N_f$. As mentioned in the main text, for general $N_f$, several fixed-point coupling values are non-zero so that different ordering channels are coupled. For large values of $N_f$ only one coupling is non-zero and a single-channel, mean-field description is possible. Out of the four stable, Lorentz invariant fixed point solutions, $\mathrm{FP}_\mathrm{V}$ does not emerge as a quantum critical point in the non Lorentz-symmetric model.}
\label{fig:qplor2}  
\end{figure*}

\begin{figure*}[h!]
\centering
\begin{subfigure}
\centering
    \includegraphics[width=0.49\linewidth]{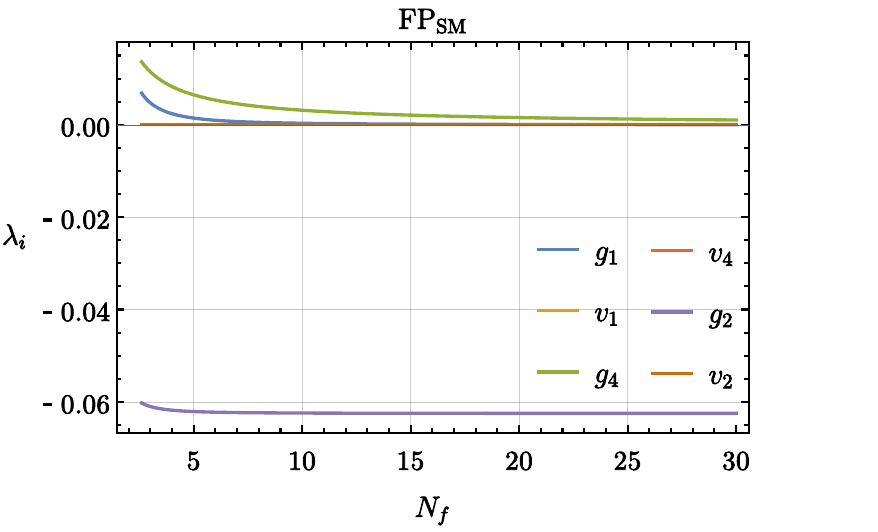}
\end{subfigure}
    \begin{subfigure}
\centering
    \includegraphics[width=0.49\linewidth]{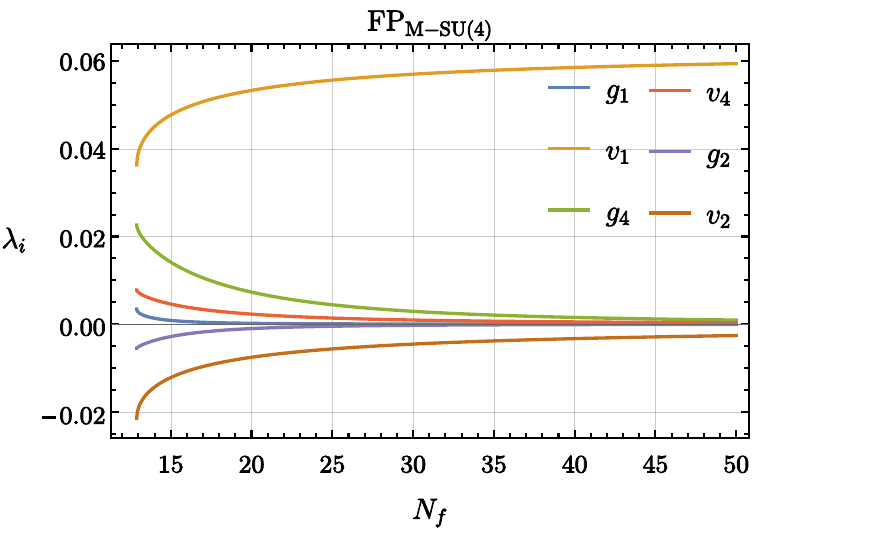}
\end{subfigure}
 \begin{subfigure}
\centering
    \includegraphics[width=0.49\linewidth]{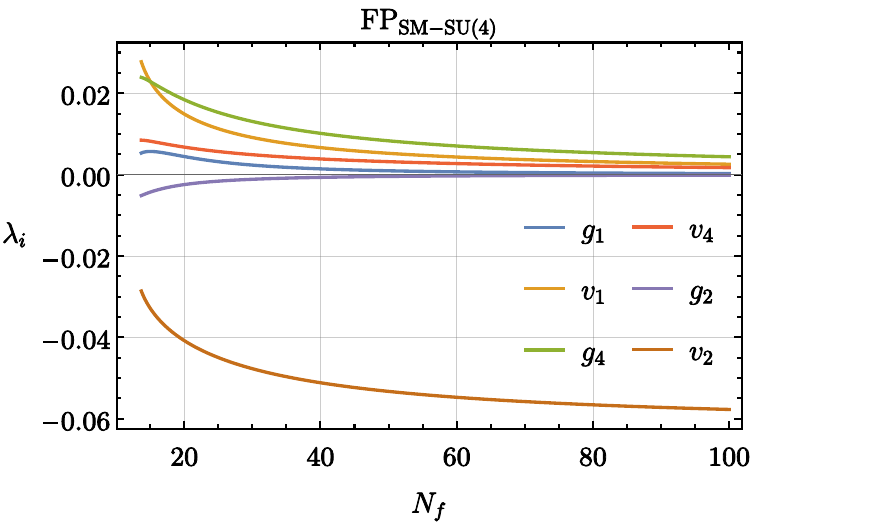}
\end{subfigure}
\begin{subfigure}
\centering
    \includegraphics[width=0.49\linewidth]{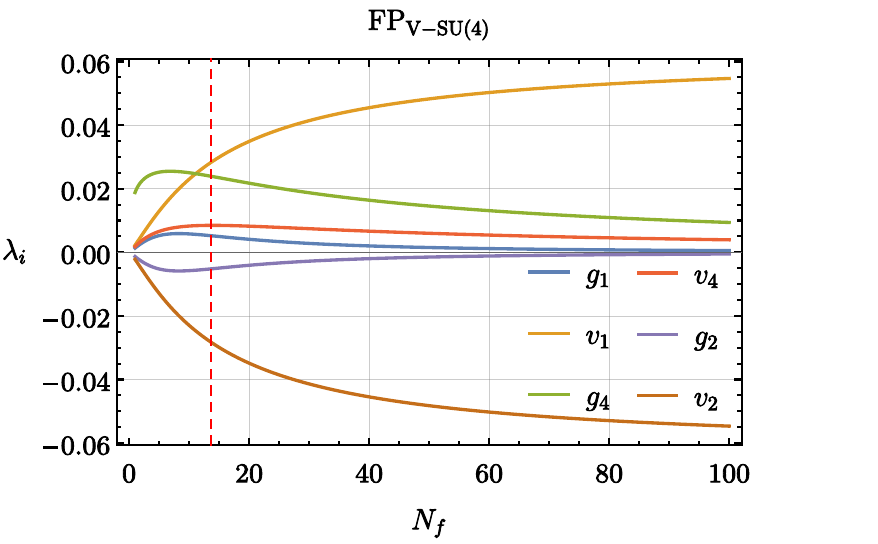}
\end{subfigure}
\begin{subfigure}
\centering
\includegraphics[width=0.49\linewidth]{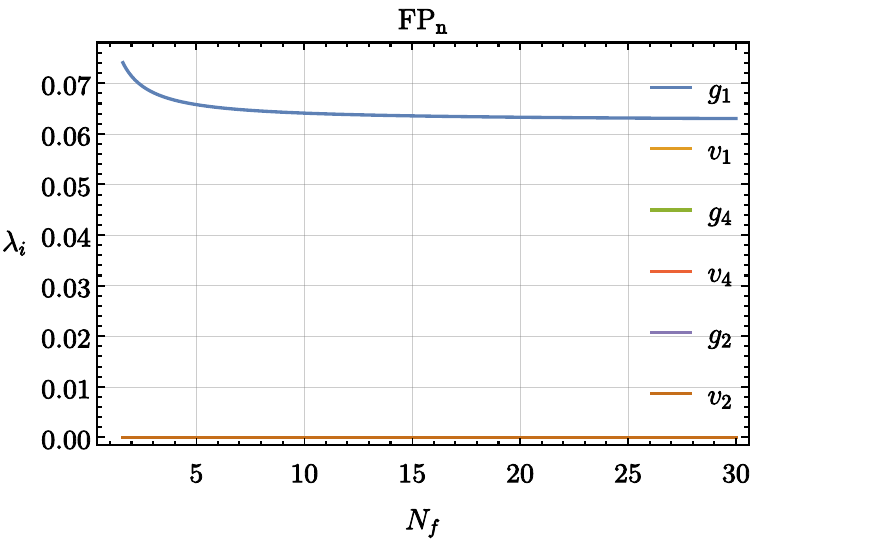}
\end{subfigure}
\begin{subfigure}    
\centering
\includegraphics[width=0.49\linewidth]{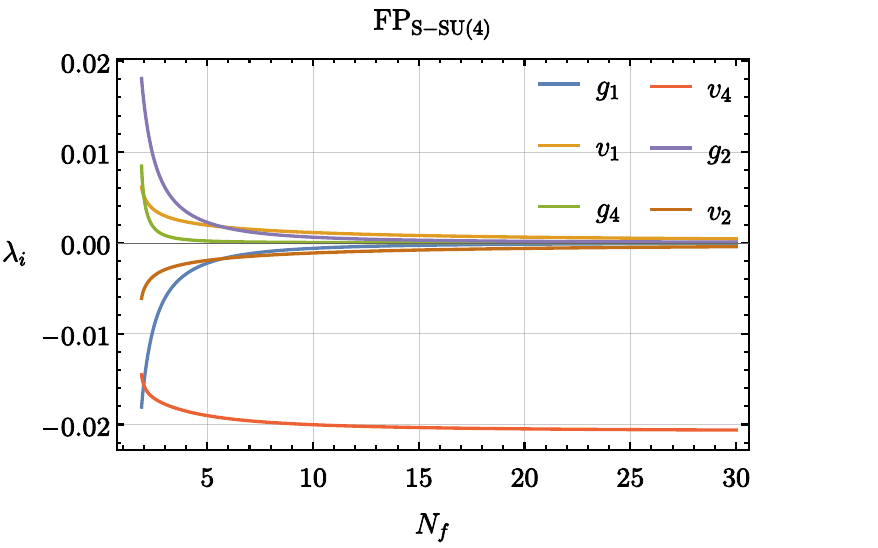}
\end{subfigure}
    \caption{Coupling values as a function of the flavor number $N_f$ of the potentially critical fixed points in the SU(4) symmetric Dirac fermion model for TBG. The red, dashed line indicates the critical value $N_f^c=13.6$ for which $\mathrm{FP}_\mathrm{V-SU(4)}$ becomes multi-critical. Fixed points FP$_i$ are associated with semi-metallic (SM), metallic SU(4) (M-SU(4)), semi-metallic SU(4), vector SU(4) (V-SU(4)), density ($n$), and scalar SU(4) (S-SU(4)) condensates.}
    \label{fig:nonlplots} 

\end{figure*}

\newpage

\section{Flow equations for the case of SU(4) symmetry relaxation}
\label{app:su4br}
We provide the beta functions for the eighteen couplings in the case of lower symmetry here
\begin{align}
    \partial_tg_1 &=g_1-\frac{4}{N_f}\biggl(g_1^2 (4 N_f-1)-g_1 (2 g_2+g_4+3 v_1^{(1)}+v_1^{(2)}+3 v_1^{(3)}+2 v_1^{(4)}+6  {v_1^{(5)}}+6  {v_2^{(1)}}+2  {v_2^{(2)}} \nonumber \\
    &+6  {v_2^{(3)}}+4  {v_2^{(4)}}+12  {v_2^{(5)}}+3  {v_4^{(1)}}+ {v_4^{(2)}}+3  {v_4^{(3)}}+2  {v_4^{(4)}}+6  {v_4^{(5)}}) \nonumber \\
    &-4 ( {g_2}  {g_4}+3  {v_2^{(1)}}  {v_4^{(1)}}+ {v_2^{(2)}}  {v_4^{(2)}}+3  {v_2^{(3)}}  {v_4^{(3)}}+2  {v_2^{(4)}}  {v_4^{(4)}}+6  {v_2^{(5)}}  {v_4^{(5)}})\biggr)\\
    \partial_t v_1^{(1)} &= v_1^{(1)}+\frac{4}{N_f} \biggl(g_1  v_1^{(1)}+2 g_2 ( v_1^{(1)}+2  v_4^{(1)})+g_4  v_1^{(1)}+4 g_4  v_2^{(1)}-4 (N_f +3) (v_1^{(1)})^2+v_1^{(1)}  v_1^{(2)}- v_1^{(1)}  v_1^{(3)} \nonumber \\ 
    &+2  v_1^{(1)}  v_1^{(4)}-2  v_1^{(1)}  v_1^{(5)}+6  v_1^{(1)}  v_2^{(1)}+2  v_1^{(1)}  v_2^{(2)}-2  v_1^{(1)}  v_2^{(3)}+4  v_1^{(1)}  v_2^{(4)}-4  v_1^{(1)}  v_2^{(5)} \nonumber \\
    &- v_1^{(1)}  v_4^{(1)}+ v_1^{(1)}  v_4^{(2)}- v_1^{(1)}  v_4^{(3)}+2  v_1^{(1)}  v_4^{(4)}-2  v_1^{(1)}  v_4^{(5)}-2  (v_1^{(3)})^2+8  v_1^{(3)}  v_2^{(3)}-4  (v_1^{(5)})^2 \nonumber \\
    &+16  v_1^{(5)}  v_2^{(5)}-4  (v_2^{(1)})^2+4  v_2^{(2)}  v_4^{(3)}-4  (v_2^{(3)})^2+4  v_2^{(3)}  v_4^{(2)}\nonumber\\
    &+8  v_2^{(4)}  v_4^{(5)}-8  (v_2^{(5)})^2+8  v_2^{(5)}  v_4^{(4)}-2  (v_4^{(1)})^2-2  (v_4^{(3)})^2-4  (v_4^{(5)})^2\biggr) \\
    \partial_t v_1^{(2)}&= v_1^{(2)} + \frac{4}{N_f}\biggl(g_1 v_1^{(2)} + g_4 v_1^{(2)} + 3 v_1^{(1)} v_1^{(2)} - (4 N_f -1)(v_1^{(2)})^2 + 
     3 v_1^{(2)} v_1^{(3)} - 2 v_1^{(2)} v_1^{(4)} - 2 (v_1^{(4)})^2 \nonumber \\ &- 6 v_1^{(2)} v_1^{(5)} - 6 (v_1^{(5)})^2 + 
     6 v_1^{(2)} v_2^{(1)} + 4 g_4 v_2^{(2)} + 2 v_1^{(2)} v_2^{(2)} + 6 v_1^{(2)} v_2^{(3)} - 4 v_1^{(2)} v_2^{(4)} + 
     8 v_1^{(4)} v_2^{(4)} - 4 (v_2^{(4)})^2 \nonumber\\
     &- 12 v_1^{(2)} v_2^{(5)} + 24 v_1^{(5)} v_2^{(5)} - 12 (v_2^{(5)})^2 + 
     3 v_1^{(2)} v_4^{(1)} \nonumber \\
     &+ 12 v_2^{(3)} v_4^{(1)} + v_1^{(2)} v_4^{(2)} + 2 g_2 (v_1^{(2)} + 2 v_4^{(2)}) + 
     3 v_1^{(2)} v_4^{(3)} + 12 v_2^{(1)} v_4^{(3)} - 2 v_1^{(2)} v_4^{(4)} - 2 (v_4^{(4)})^2 - 6 v_1^{(2)} v_4^{(5)} - 
     6 (v_4^{(5)})^2\biggr)\\
    \partial v_1^{(3)} &= v_1^{3} + \frac{4}{N_f} \biggl(g_1 v_1^{(3)} + g_4 v_1^{(3)} - 5 v_1^{(1)} v_1^{(3)} + v_1^{(2)} v_1^{(3)}  - 
     (4 N_f+1) (v_1^{(3)})^2 - 2 v_1^{(3)} v_1^{(4)} + 2 v_1^{(3)} v_1^{(5)} \nonumber \\ 
     &- 4 v_1^{(4)} v_1^{(5)} + 6 v_1^{(3)} v_2^{(1)} + 
     2 v_1^{(3)} v_2^{(2)} + 4 g_4 v_2^{(3)} + 8 v_1^{(1)} v_2^{(3)} - 2 v_1^{(3)} v_2^{(3)} - 8 v_2^{(1)} v_2^{(3)} - 
     4 v_1^{(3)} v_2^{(4)}\nonumber\\
     &+ 8 v_1^{(5)} v_2^{(4)} + 4 v_1^{(3)} v_2^{(5)} + 8 v_1^{(4)} v_2^{(5)} - 8 v_2^{(4)} v_2^{(5)} - 
     v_1^{(3)} v_4^{(1)} + 4 v_2^{(2)} v_4^{(1)} + v_1^{(3)} v_4^{(2)} + 4 v_2^{(1)} v_4^{(2)} \nonumber \\
     &- v_1^{(3)} v_4^{(3)} - 
     4 v_4^{(1)} v_4^{(3)} + 2 g_2 (v_1^{(3)} + 2 v_4^{(3)}) - 2 v_1^{(3)} v_4^{(4)} + 2 v_1^{(3)} v_4^{(5)} + 
     16 v_2^{(5)} v_4^{(5)} - 4 v_4^{(4)} v_4^{(5)}\biggr) \\
    \partial_t v_1^{(4)}&=v_1^{(4)} + \frac{4}{N_f}\biggl(g_1 v_1^{(4)} + g_4 v_1^{(4)} + 3 v_1^{(1)} v_1^{(4)} - 3 v_1^{(2)} v_1^{(4)} - 3 v_1^{(3)} v_1^{(4)} - 
    4 N_f (v_1^{(4)})^2 - 6 v_1^{(3)} v_1^{(5)}  \nonumber \\ 
    &+ 6 v_1^{(4)} v_2^{(1)} + 2 v_1^{(4)} v_2^{(2)} - 6 v_1^{(4)} v_2^{(3)} + 
    12 v_1^{(5)} v_2^{(3)} + 4 g_4 v_2^{(4)} + 4 v_1^{(2)} v_2^{(4)} - 4 v_2^{(2)} v_2^{(4)} \nonumber \\
    &+ 12 v_1^{(3)} v_2^{(5)} - 
    12 v_2^{(3)} v_2^{(5)} + 3 v_1^{(4)} v_4^{(1)} + 12 v_2^{(5)} v_4^{(1)} - v_1^{(4)} v_4^{(2)} - 3 v_1^{(4)} v_4^{(3)} - 
    2 v_4^{(2)} v_4^{(4)} \nonumber \\ 
    &+ 2 g_2 (v_1^{(4)} + 2 v_4^{(4)}) + 12 v_2^{(1)} v_4^{(5)} - 6 v_4^{(3)} v_4^{(5)}\biggr)\\
    \partial_t v_1^{(5)} &= v_1^{(5)} + \frac{4}{N_f}\biggl(g_1 v_1^{(5)} + 2 g_2 v_1^{(5)} + g_4 v_1^{(5)} - 5 v_1^{(1)} v_1^{(5)} - 3 v_1^{(2)} v_1^{(5)} - 
    4 N_f (v_1^{(5)})^2 + 6 v_1^{(5)} \nonumber \\ 
    & + 4 v_1^{(4)} v_2^{(3)} + 2 v_1^{(5)} v_2^{(3)} - 
    4 v_2^{(3)} v_2^{(4)} + v_1^{(3)} (-2 v_1^{(4)} + v_1^{(5)} + 4 v_2^{(4)}) + 4 g_4 v_2^{(5)} + 8 v_1^{(1)} v_2^{(5)} \nonumber \\ 
    &+ 4 v_1^{(2)} v_2^{(5)} - 8 v_2^{(1)} v_2^{(5)} - 4 v_2^{(2)} v_2^{(5)} - v_1^{(5)} v_4^{(1)} 
    + 4 v_2^{(4)} v_4^{(1)} - 
    v_1^{(5)} v_4^{(2)} 
    + v_1^{(5)} v_4^{(3)} \nonumber \\ 
    &+ 8 v_2^{(5)} v_4^{(3)} + 4 v_2^{(1)} v_4^{(4)} - 2 v_4^{(3)} v_4^{(4)} + 
    4 g_2 v_4^{(5)} + 8 v_2^{(3)} v_4^{(5)} - 4 v_4^{(1)} v_4^{(5)} - 2 v_4^{(2)} v_4^{(5)}\biggr) \\
    \partial_t g_4&= g_4 + \frac{4}{N_f}  \biggl(4 g_1 g_2 - 2 g_2^2 - 3 g_1 g_4 + 6 g_2 g_4 - 3 g_4^2 + 12 g_4^2 N_f - 
  9 g_4 v_1^{(1)} - 3 g_4 v_1^{(2)} \nonumber\\ 
  &- 9 g_4 v_1^{(3)} - 6 g_4 v_1^{(4)} - 18 g_4 v_1^{(5)} + 18 g_4 v_2^{(1)} + 
  12 v_1^{(1)} v_2^{(1)} - 6 (v_2^{(1)})^2 + 6 g_4 v_2^{(2)} \nonumber \\ 
  &+ 4 v_1^{(2)} v_2^{(2)} - 2 (v_2^{(2)})^2 + 18 g_4 v_2^{(3)} + 
  12 v_1^{(3)} v_2^{(3)} - 6 (v_2^{(3)})^2 + 12 g_4 v_2^{(4)} + 8 v_1^{(4)} v_2^{(4)} \nonumber \\
  &- 4 (v_2^{(4)})^2 + 
  36 g_4 v_2^{(5)} + 24 v_1^{(5)} v_2^{(5)} - 12 (v_2^{(5)})^2 - 9 g_4 v_4^{(1)} - 3 g_4 v_4^{(2)} - 
  9 g_4 v_4^{(3)} - 6 g_4 v_4^{(4)} - 18 g_4 v_4^{(5)}\biggr) \\
    \partial_t v_4^{(1)}&= v_4^{(1)} + \frac{4}{N_f}  \biggl(4 g_1 v_2^{(1)} + 4 v_1^{(3)} v_2^{(2)} + 4 v_1^{(2)} v_2^{(3)} - 4 v_2^{(2)} v_2^{(3)} + 8 v_1^{(5)} v_2^{(4)} + 
  8 v_1^{(4)} v_2^{(5)} - 8 v_2^{(4)} v_2^{(5)} - 3 g_1 v_4^{(1)} \nonumber \\
  &- 3 g_4 v_4^{(1)} - v_1^{(1)} v_4^{(1)} - 3 v_1^{(2)} v_4^{(1)} + 
  3 v_1^{(3)} v_4^{(1)} - 6 v_1^{(4)} v_4^{(1)} + 6 v_1^{(5)} v_4^{(1)} + 2 v_2^{(1)} v_4^{(1)} + 6 v_2^{(2)} v_4^{(1)} - 
  6 v_2^{(3)} v_4^{(1)} \nonumber \\ 
  &+ 12 v_2^{(4)} v_4^{(1)} - 12 v_2^{(5)} v_4^{(1)} + 3 (v_4^{(1)})^2 + 12 N_f (v_4^{(1)})^2 + 
  g_2 (4 v_1^{(1)} - 4 v_2^{(1)} + 6 v_4^{(1)}) - 3 v_4^{(1)} v_4^{(2)} - 4 v_1^{(3)} v_4^{(3)} \nonumber \\ 
  &+ 8 v_2^{(3)} v_4^{(3)} + 
  3 v_4^{(1)} v_4^{(3)} - 6 v_4^{(1)} v_4^{(4)} - 8 v_1^{(5)} v_4^{(5)} + 16 v_2^{(5)} v_4^{(5)} + 6 v_4^{(1)} v_4^{(5)}\biggr) \\
    \partial_t v_4^{(2)}&= v_4^{(2)} + \frac{4}{N_f}  \biggl(12 v_1^{(3)} v_2^{(1)} + 4 g_1 v_2^{(2)} + 12 v_1^{(1)} v_2^{(3)} - 12 v_2^{(1)} v_2^{(3)} - 3 g_1 v_4^{(2)} - 
   3 g_4 v_4^{(2)} - 9 v_1^{(1)} v_4^{(2)} \nonumber \\
   &- 3 v_1^{(2)} v_4^{(2)} - 9 v_1^{(3)} v_4^{(2)} + 6 v_1^{(4)} v_4^{(2)} + 
   18 v_1^{(5)} v_4^{(2)} + 18 v_2^{(1)} v_4^{(2)} + 6 v_2^{(2)} v_4^{(2)} + 18 v_2^{(3)} v_4^{(2)} \nonumber \\
   &- 12 v_2^{(4)} v_4^{(2)} - 
   36 v_2^{(5)} v_4^{(2)} - 9 v_4^{(1)} v_4^{(2)} - 3 (v_4^{(2)})^2 (1- 4 N_f) + 
   g_2 (4 v_1^{(2)} - 4 v_2^{(2)} + 6 v_4^{(2)}) \nonumber \\ &
   - 9 v_4^{(2)} v_4^{(3)} - 4 v_1^{(4)} v_4^{(4)} + 8 v_2^{(4)} v_4^{(4)} + 
   6 v_4^{(2)} v_4^{(4)} - 12 v_1^{(5)} v_4^{(5)} + 24 v_2^{(5)} v_4^{(5)} + 18 v_4^{(2)} v_4^{(5)}\biggr) \\
    \partial_t v_4^{(3)}&= v_4^{(3)} + \frac{4}{N_f} \biggl(4 v_1^{(2)} v_2^{(1)} + 4 v_1^{(1)} v_2^{(2)} - 4 v_2^{(1)} v_2^{(2)} + 4 g_1 v_2^{(3)} + 16 v_1^{(5)} v_2^{(5)} - 
  8 (v_2^{(5)})^2 - 4 v_1^{(3)} v_4^{(1)} \nonumber \\
  &+ 8 v_2^{(3)} v_4^{(1)} - 3 g_1 v_4^{(3)} - 3 g_4 v_4^{(3)} - v_1^{(1)} v_4^{(3)} - 
  3 v_1^{(2)} v_4^{(3)} + 3 v_1^{(3)} v_4^{(3)} + 6 v_1^{(4)} v_4^{(3)} \nonumber \\ 
  &- 6 v_1^{(5)} v_4^{(3)} + 2 v_2^{(1)} v_4^{(3)} + 
  6 v_2^{(2)} v_4^{(3)} - 6 v_2^{(3)} v_4^{(3)} - 12 v_2^{(4)} v_4^{(3)} + 12 v_2^{(5)} v_4^{(3)} + 3 v_4^{(1)} v_4^{(3)} - 
  3 v_4^{(2)} v_4^{(3)} + 3 (v_4^{(3)})^2  \nonumber \\
  &+12 N_f (v_4^{(3)})^2 + g_2 (4 v_1^{(3)} - 4 v_2^{(3)} + 6 v_4^{(3)}) - 
  4 v_1^{(5)} v_4^{(4)} + 8 v_2^{(5)} v_4^{(4)} + 6 v_4^{(3)} v_4^{(4)} - 4 v_1^{(4)} v_4^{(5)} + 8 v_2^{(4)} v_4^{(5)} - 
  6 v_4^{(3)} v_4^{(5)}\biggr) \\
    \partial_t v_4^{(4)}&= v_4^{(4)} + \frac{4}{N_f}   \biggl(12 v_1^{(5)} v_2^{(1)} + 4 g_1 v_2^{(4)} + 12 v_1^{(1)} v_2^{(5)} - 12 v_2^{(1)} v_2^{(5)} - 2 v_1^{(4)} v_4^{(2)} + 
  4 v_2^{(4)} v_4^{(2)} - 6 v_1^{(5)} v_4^{(3)} \nonumber \\ 
  &+ 12 v_2^{(5)} v_4^{(3)} - 3 g_1 v_4^{(4)} - 3 g_4 v_4^{(4)} - 
  9 v_1^{(1)} v_4^{(4)} + v_1^{(2)} v_4^{(4)} + 9 v_1^{(3)} v_4^{(4)} + 18 v_2^{(1)} v_4^{(4)} - 2 v_2^{(2)} v_4^{(4)} \nonumber\\
  &-18 v_2^{(3)} v_4^{(4)} - 9 v_4^{(1)} v_4^{(4)} + 3 v_4^{(2)} v_4^{(4)} + 9 v_4^{(3)} v_4^{(4)} + 12 N_f (v_4^{(4)})^2 + 
  g_2 (4 v_1^{(4)} - 4 v_2^{(4)} + 6 v_4^{(4)}) \nonumber \\
  &- 6 v_1^{(3)} v_4^{(5)} + 12 v_2^{(3)} v_4^{(5)}\biggr) \\
    \partial_t v_4^{(5)}&= v_4^{(5)} + \frac{4}{N_f} \biggl(4 v_1^{(4)} v_2^{(1)} + 8 v_1^{(5)} v_2^{(3)} + 4 v_1^{(1)} v_2^{(4)} -4 v_2^{(1)} v_2^{(4)} + 4 g_1 v_2^{(5)} + 
  8 v_1^{(3)} v_2^{(5)} - 8 v_2^{(3)} v_2^{(5)} 
  - 4 v_1^{(5)} v_4^{(1)} + 8 v_2^{(5)} v_4^{(1)} \nonumber\\ &- 2 v_1^{(5)} v_4^{(2)} + 
  4 v_2^{(5)} v_4^{(2)} - 2 v_1^{(4)} v_4^{(3)} + 4 v_2^{(4)} v_4^{(3)} - 2 v_1^{(3)} v_4^{(4)} + 4 v_2^{(3)} v_4^{(4)} - 
  3 g_1 v_4^{(5)} - 3 g_4 v_4^{(5)} - v_1^{(1)} v_4^{(5)} + v_1^{(2)} v_4^{(5)} \nonumber\\
  &- 3 v_1^{(3)} v_4^{(5)} + 2 v_2^{(1)} v_4^{(5)} - 
  2 v_2^{(2)} v_4^{(5)} + 6 v_2^{(3)} v_4^{(5)} + 3 v_4^{(1)} v_4^{(5)} + 3 v_4^{(2)} v_4^{(5)} - 3 v_4^{(3)} v_4^{(5)} \nonumber\\
  &+ 
  12 N_f (v_4^{(5)})^2 + g_2 \bigl(4 v_1^{(5)} - 4 v_2^{(5)} + 6 v_4^{(5)}\bigr)\biggr) \\
    \partial_t g_2 &= g_2 - \frac{4}{N_f}  \biggr(g_1 (g_2 - 2 g_4) - 4 g_2^2 N_f + 
  g_2 (g_4 + 3 v_1^{(1)} + v_1^{(2)} + 3 v_1^{(3)} + 2 v_1^{(4)} + 6 v_1^{(5)} - 3 v_4^{(1)} - v_4^{(2)} - 
     3 v_4^{(3)} - 2 v_4^{(4)} - 6 v_4^{(5)}) \nonumber\\
  &+ 2 (-3 v_1^{(1)} v_4^{(1)} + 3 v_2^{(1)} v_4^{(1)} - v_1^{(2)} v_4^{(2)} + v_2^{(2)} v_4^{(2)} - 3 v_1^{(3)} v_4^{(3)} + 
     3 v_2^{(3)} v_4^{(3)} \nonumber\\
  &- 2 v_1^{(4)} v_4^{(4)} + 2 v_2^{(4)} v_4^{(4)} - 6 v_1^{(5)} v_4^{(5)} + 6 v_2^{(5)} v_4^{(5)})\biggl) \\
    \partial_t v_2^{(1)} &= v_2^{(1)} + \frac{4}{N_f} \biggr(2 g_4 v_1^{(1)} + 2 (v_1^{(1)})^2 + 2 (v_1^{(3)})^2 + 4 (v_1^{(5)})^2 - g_1 v_2^{(1)} - g_4 v_2^{(1)} - 
  3 v_1^{(1)} v_2^{(1)} - v_1^{(2)} v_2^{(1)} + v_1^{(3)} v_2^{(1)} \nonumber\\
  &- 2 v_1^{(4)} v_2^{(1)} + 2 v_1^{(5)} v_2^{(1)} + 
  4  (v_2^{(1)})^2 (N_f+2)- 4 v_1^{(3)} v_2^{(3)} + 8 (v_2^{(3)})^2 - 8 v_1^{(5)} v_2^{(5)} + 16 (v_2^{(5)})^2 + 
  2 g_1 v_4^{(1)} \nonumber\\
  &- 2 g_2 v_4^{(1)} - v_2^{(1)} v_4^{(1)} + 2 (v_4^{(1)})^2 + 2 v_1^{(3)} v_4^{(2)} + v_2^{(1)} v_4^{(2)} - 
  2 v_2^{(3)} v_4^{(2)} + 2 v_1^{(2)} v_4^{(3)} - v_2^{(1)} v_4^{(3)} - 2 v_2^{(2)} v_4^{(3)} + 2 (v_4^{(3)})^2 \nonumber\\
  &+ 4 v_1^{(5)} v_4^{(4)} + 
  2 v_2^{(1)} v_4^{(4)} - 4 v_2^{(5)} v_4^{(4)} + 4 v_1^{(4)} v_4^{(5)} - 2 v_2^{(1)} v_4^{(5)} - 4 v_2^{(4)} v_4^{(5)} + 4 (v_4^{(5)})^2\biggl) \\ 
    \partial_t v_2^{(2)} &= v_2^{(2)} + \frac{4}{N_f} \biggl(2 g_4 v_1^{(2)} + 2 (v_1^{(4)})^2 + 6 (v_1^{(5)})^2 - g_1 v_2^{(2)} - g_4 v_2^{(2)} - 3 v_1^{(1)} v_2^{(2)} - 
  v_1^{(2)} v_2^{(2)} - 3 v_1^{(3)} v_2^{(2)} + 2 v_1^{(4)} v_2^{(2)} \nonumber\\
  &+ 6 v_1^{(5)} v_2^{(2)} + 4 N_f (v_2^{(2)})^2 - 
  4 v_1^{(4)} v_2^{(4)} + 8 (v_2^{(4)})^2 - 12 v_1^{(5)} v_2^{(5)} + 24 (v_2^{(5)})^2 + 6 v_1^{(3)} v_4^{(1)} + 
  3 v_2^{(2)} v_4^{(1)} - 6 v_2^{(3)} v_4^{(1)} \nonumber\\
  &+ 2 g_1 v_4^{(2)} - 2 g_2 v_4^{(2)} + v_2^{(2)} v_4^{(2)} + 6 v_1^{(1)} v_4^{(3)} - 
  6 v_2^{(1)} v_4^{(3)} + 3 v_2^{(2)} v_4^{(3)} - 2 v_2^{(2)} v_4^{(4)} + 2 (v_4^{(4)})^2 - 6 v_2^{(2)} v_4^{(5)} + 6 (v_4^{(5)})^2\biggr)\\
    \partial_t v_2^{(3)} &= v_2^{(3)} - \frac{4}{N_f}  \biggl(-4 v_1^{(4)} v_1^{(5)} + 4 v_1^{(3)} v_2^{(1)} + g_1 v_2^{(3)} + v_1^{(2)} v_2^{(3)} - v_1^{(3)} v_2^{(3)} - 2 v_1^{(4)} v_2^{(3)} + 
  2 v_1^{(5)} v_2^{(3)} - 16 v_2^{(1)} v_2^{(3)} \nonumber\\
  &- 4 N_f (v_2^{(3)})^2 + g_4 (-2 v_1^{(3)} + v_2^{(3)}) + 
  4 v_1^{(5)} v_2^{(4)} + 4 v_1^{(4)} v_2^{(5)} - 16 v_2^{(4)} v_2^{(5)} - 2 v_1^{(2)} v_4^{(1)} + 2 v_2^{(2)} v_4^{(1)} + 
  v_2^{(3)} v_4^{(1)} \nonumber\\
  &+ v_1^{(1)} (-4 v_1^{(3)} + 3 v_2^{(3)} - 2 v_4^{(2)}) + 2 v_2^{(1)} v_4^{(2)} - v_2^{(3)} v_4^{(2)} - 
  2 g_1 v_4^{(3)} + 2 g_2 v_4^{(3)} + v_2^{(3)} v_4^{(3)} - 4 v_4^{(1)} v_4^{(3)} + 2 v_2^{(3)} v_4^{(4)} \nonumber\\
  &- 8 v_1^{(5)} v_4^{(5)} - 
  2 v_2^{(3)} v_4^{(5)} + 8 v_2^{(5)} v_4^{(5)} - 4 v_4^{(4)} v_4^{(5)}\biggr)\\
    \partial_t v_2^{(4)} &= v_2^{(4)} - \frac{4}{N_f} \biggl(-6 v_1^{(3)} v_1^{(5)} + 2 v_1^{(4)} v_2^{(2)} + 6 v_1^{(5)} v_2^{(3)} + g_1 v_2^{(4)} + 3 v_1^{(1)} v_2^{(4)} - 
  3 v_1^{(3)} v_2^{(4)} - 8 v_2^{(2)} v_2^{(4)} - 4 N_f (v_2^{(4)})^2 \nonumber\\
  &+ g_4 (-2 v_1^{(4)} + v_2^{(4)}) + 
  v_1^{(2)} (-2 v_1^{(4)} + v_2^{(4)}) + 6 v_1^{(3)} v_2^{(5)} - 24 v_2^{(3)} v_2^{(5)} - 6 v_1^{(5)} v_4^{(1)} - 
  3 v_2^{(4)} v_4^{(1)} + 6 v_2^{(5)} v_4^{(1)} \nonumber\\
  &+ v_2^{(4)} v_4^{(2)} + 3 v_2^{(4)} v_4^{(3)} - 2 g_1 v_4^{(4)} + 2 g_2 v_4^{(4)} - 
  2 v_4^{(2)} v_4^{(4)} - 6 v_1^{(1)} v_4^{(5)} + 6 v_2^{(1)} v_4^{(5)} - 6 v_4^{(3)} v_4^{(5)}\biggr) \\
    \partial_t v_2^{(5)} &= v_2^{(5)} - \frac{4}{N_f}  \biggl(-4 v_1^{(1)} v_1^{(5)} - 2 v_1^{(2)} v_1^{(5)} + 4 v_1^{(5)} v_2^{(1)} + 2 v_1^{(5)} v_2^{(2)} + 2 v_1^{(4)} v_2^{(3)} - 
  8 v_2^{(3)} v_2^{(4)} + g_1 v_2^{(5)} + 3 v_1^{(1)} v_2^{(5)} \nonumber\\
  &+ v_1^{(2)} v_2^{(5)} - 16 v_2^{(1)} v_2^{(5)} - 8 v_2^{(2)} v_2^{(5)} - 
  4 N_f (v_2^{(5)})^2 + g_4 (-2 v_1^{(5)} + v_2^{(5)}) - 2 v_1^{(4)} v_4^{(1)} + 2 v_2^{(4)} v_4^{(1)} \nonumber\\
  &+ v_2^{(5)} v_4^{(1)} + 
  v_2^{(5)} v_4^{(2)} - 4 v_1^{(5)} v_4^{(3)} + 3 v_2^{(5)} v_4^{(3)} - 2 v_1^{(1)} v_4^{(4)} + 2 v_2^{(1)} v_4^{(4)} - 
  2 v_4^{(3)} v_4^{(4)} + v_1^{(3)} (-2 v_1^{(4)} + 2 v_2^{(4)} + v_2^{(5)} - 4 v_4^{(5)}) \nonumber\\
  &- 2 g_1 v_4^{(5)} + 
  2 g_2 v_4^{(5)} + 4 v_2^{(3)} v_4^{(5)} - 4 v_4^{(1)} v_4^{(5)} - 2 v_4^{(2)} v_4^{(5)}\biggr)
\end{align}

\end{widetext}

\end{appendix}
\end{document}